# A Cognitive Agentic AI Framework for Interactive Process Analysis and Optimization

Trung Ngoc Trinh[a,*], Roman Tschentscher[a], Trong Dong Nguyen[b], Trung Thanh Nguyen[c], Anna Trubetskaya[d],

[a]SINTEF Industry, 7034 Trondheim, Norway

[b]Department of Marine Technology, Norwegian University of Science and Technology, Trondheim, Norway

[c]Faculty of Information Technology, Da Nang University of Science and Technology, Vietnam

[d]Faculty of Biosciences and Aquaculture, Nord University, Steinkjer, Norway

* Corresponding author: Trung Ngoc Trinh; trung.trinh@sintef.no

## Abstract

Complex multivariable process analysis and optimization require the coordinated use of process simulations, structured data, machine-learning (ML) models, optimization algorithms, and scientific literature, which are often accessed through separate software environments. This study presents a Cognitive Agentic AI Framework (CAAF) that enables engineers to conduct interactive and iterative process investigations through natural-language prompts within an established process flowsheet. Developed following harness-engineering principles, CAAF separates LLM-based task interpretation, planning, knowledge synthesis, and context management from numerical calculations performed by specialized engineering tools. Dedicated Providers connect structured databases, scientific-document retrieval, ML-assisted Bayesian optimization (BO), and COCO process simulation. The framework was demonstrated through the analysis and optimization of a methyl acetate production process. Through successive prompts, an engineer investigated multivariable process relationships, performed multi-objective BO over 4000 iterations to maximize carbon conversion yield (C-Yield) and minimize specific energy cost (C-SEC), refined the operating domain based on engineering insights, and requested process simulation verification of selected optimization candidates. Within the restricted operating domain, re-evaluation of 16 selected candidates using process simulation demonstrated close agreement with ML-based predictions, achieving a mean absolute error of 0.0085 (equivalent to 0.85% in C-Yield). The highest C-Yield achieved through BO and subsequently verified by COCO process simulation was 0.921, compared with a maximum of approximately 0.81 in the initial simulation dataset, representing an improvement of about 10%. The framework also incorporates mechanisms to report non-converged simulations, exclude invalid outputs, and continue subsequent evaluations. These findings demonstrate how human-guided interaction can effectively coordinate complementary engineering resources, support iterative process analysis and optimization, and integrate surrogate-based optimization with rigorous simulation-based verification within a fixed process topology.

## 1. Introduction

Process analysis and optimization require engineers to connect process understanding with quantitative evidence. Even when the process simulation flowsheet is well established, interactions among reaction conditions, separation performance, recycle streams, and equipment parameters can create complex, nonlinear relationships between process inputs and competing performance objectives [1–4]. Understanding these interactions is essential for gaining meaningful process insights and enabling effective global optimization of complex multivariable processes. Investigating these relationships commonly involves process simulation, structured-data analysis, optimization algorithms, and scientific literature [2–6]. An efficient strategy for complex multivariable process optimization is to integrate rigorous process simulation with machine-learning (ML) surrogate models and Bayesian optimization (BO). Simulation-generated datasets are used to train ML models that approximate nonlinear relationships between process inputs and performance objectives. These computationally efficient surrogates subsequently serve as the evaluation engine for BO, enabling extensive global search and multi-objective optimization without repeatedly executing computationally expensive process simulations. The strategies have been demonstrated in previous studies, including ML-assisted global flowsheet optimization, surrogate-based Bayesian optimization of chemical processes and process-wide optimization integrating Aspen HYSYS with artificial neural networks [2–4,6–11] Selected high-performing solutions are subsequently re-evaluated using rigorous process simulation to verify predicted performance and assess their consistency with the reference process model [8]. By combining the computational efficiency of ML-based optimization with targeted simulation-based verification, this strategy enables the identification of promising near-global solutions while substantially reducing computational time and cost, particularly when trained surrogate models are reused across multiple optimization investigations.

Experienced process engineers possess extensive domain knowledge of process behavior, operating constraints, and performance optimization. However, investigating complex multivariable processes with nonlinear interactions often exceeds conventional engineering intuition and requires diverse technical skills, including data analytics, process simulation, machine learning, optimization, and programming [11,12]. These requirements can limit the accessibility and efficiency of advanced computational methods, even for engineers with strong domain expertise [13]. Large Language Model (LLM)-powered Agentic AI offers a promising solution by connecting engineering knowledge with specialized computational resources through natural-language interaction [11,13,14]. By coordinating multiple agents and tools, engineers can analyze process data, execute simulations, perform optimization, and interpret results using domain-oriented prompts without directly operating individual software tools or writing extensive code. Furthermore, AI System with LLM maintains context across successive interactions, enabling engineers to refine investigation objectives and optimization constraints based on intermediate findings while retaining control over engineering decisions. Recent studies have demonstrated LLM-based interaction with AVEVA Process Simulation [13] and Aspen Plus [11], as well as multi-agent approaches for interpreting process-control policies [15]. Broader applications, including ChemCrow and Coscientist, demonstrate the integration of LLMs with scientific tools and computational workflows [16,17]. Recent reviews highlight the potential of LLM-based multi-agent systems in chemical engineering to coordinate specialized agents, integrate heterogeneous engineering data and computational tools, and support complex engineering workflows, while emphasizing challenges related to reliability, transparency, and system integration [14]. These developments motivate the integration of Agentic AI, process analysis, ML-assisted Bayesian optimization, and simulation-based verification into a unified interactive engineering framework.

Despite the potential of LLM-powered Agentic AI, relying heavily on LLM reasoning for task abstraction, planning, and coordination across heterogeneous engineering data, documents, and computational models can increase the risk of hallucinations, inconsistent decisions, and incorrect tool execution [18,19]. These limitations are particularly critical in process engineering, where numerical results and engineering conclusions must remain consistent with physical principles and process constraints.

Recent developments in harness engineering seek to improve the reliability of LLM-powered agents by establishing a structured execution environment that integrates agent orchestration, controlled tool interfaces, domain-specific knowledge, memory, and verification mechanisms [20–24]. Rather than relying on unrestricted LLM reasoning, this approach separates cognitive reasoning and task coordination from numerical calculations performed by specialized engineering tools, while supporting adaptive investigation workflows. These principles provide the architectural foundation for the proposed CAAF, enabling interactive process analysis, ML-assisted Bayesian optimization, and rigorous simulation-based verification within a controlled and traceable engineering environment.

Another important aspect of Agentic AI in process engineering is the integration of the Human-in-the-Loop (HITL) paradigm, which ensures that engineering expertise remains central to AI-assisted investigations [25–28]. Rather than replacing human expertise, Agentic AI supports engineers in performing complex analytical and optimization tasks while retaining human oversight of engineering decisions. Through natural-language interaction, engineers define investigation objectives and constraints, evaluate intermediate findings, and refine subsequent analyses based on their domain knowledge. In particular, engineers can interpret process insights and optimization results, adjust the search space, and request simulation-based verification of selected solutions. This iterative interaction between human expertise and AI-assisted computational capabilities enables a flexible and interactive investigation workflow, where Agentic AI coordinates engineering resources while engineers retain responsibility for interpreting results and making final decisions.

Building on these developments, this study presents a Cognitive Agentic AI Framework (CAAF) for interactive process analysis and optimization within a fixed process flowsheet. Rather than automating individual engineering tasks independently, CAAF establishes a unified, multi-turn investigation environment in which engineers can formulate research questions, explore process behavior, interpret intermediate findings, and progressively refine optimization objectives and constraints through natural-language interactions. Developed following harness-engineering principles, the framework coordinates heterogeneous engineering knowledge and computational resources while maintaining human expertise and domain-specific verification throughout the investigation.

The main contributions of this work are summarized as follows:

- An interactive Cognitive Agentic AI framework that integrates cognitive reasoning, investigation planning, and engineering tool orchestration to support large-scale process data analysis and complex multivariable optimization. The framework enables engineers to explore nonlinear process interactions, interpret optimization outcomes, and iteratively guide subsequent investigations based on engineering insights.
- A harness-engineering architecture that combines ontology-guided reasoning, engineering memory, federated knowledge access, specialized agents, and tool-based verification. By separating LLM-driven reasoning and coordination from numerical computations performed by domain-specific tools, the architecture provides a structured and traceable investigation environment, including mechanisms for handling non-converged simulations.
- A methyl acetate production case study that demonstrates the practical application of CAAF by integrating structured process data, scientific literature, ML-assisted multi-objective Bayesian optimization, and rigorous COCO process simulation. Through successive engineering prompts, the case illustrates a continuous investigation process from multivariable analysis and optimization to human-guided refinement of the search space and simulation-based verification of selected solutions.

# 2. Methodology

## 2.1. Overall Concept

CAAF connects the engineer to the resources required for an iterative process investigation, as illustrated in Figure 1. The engineer expresses an objective or question in natural language. The framework interprets the request, coordinates relevant computational tasks, and returns numerical results, visualizations, and supporting explanations. The engineer can then refine the objective, examine an alternative interpretation, modify parameter bounds, or request verification. These interactions maintain continuity between the initial engineering question and the evidence obtained in subsequent calculations.

The workflow distinguishes between resource preparation and interactive use. During preparation, an established process simulation flowsheet is evaluated across a defined parameter domain to generate process data. Outputs from successfully converged simulations are organized into a structured database, while a selected subset of these data is used to train ML surrogate models that capture the nonlinear relationships between process parameters and performance objectives required for optimization. Scientific documents are separately prepared for retrieval. This stage produces the simulator, database, trained models, optimization routines, and document resource available to the agents. During interactive use, CAAF invokes these prepared resources and requests additional process simulation evaluations when the investigation requires them.

Flowsheet topology is fixed throughout data generation, analysis, optimization, and candidate verification. The study varies selected operating and equipment parameters within this configuration. The reference simulation therefore provides a common basis for the stored data, surrogate predictions, and re-evaluated candidates. In the optimization workflow, the trained ML model acts as the numerical prediction engine, and Bayesian optimization searches the specified parameter ranges according to the requested objectives. The engineer can inspect the resulting trade-offs and select conditions for process simulation re-evaluation, then use the comparison to guide the next interaction. CAAF incorporates mechanisms to detect and report simulation convergence failures and execution errors, enabling engineers to monitor the simulation-based evaluation process and identify unsuccessful simulation runs.

CAAF adopts a federated resource organization in which providers expose the capabilities of separately managed databases, simulation models, optimization tools, and document repositories. The providers translate structured requests into resource-specific operations and return results to the coordinating agents. This arrangement allows participating groups to maintain their engineering resources within their respective environments. In the case study, four partner roles represent the complementary resources required for the investigation. The division of responsibilities separates preparation and numerical execution from natural-language coordination and interpretation.

The human-in-the-loop workflow operates through successive requests and reviews of returned evidence. An initial analysis may suggest a promising region, an optimization result may reveal a trade-off, and a simulator comparison may motivate additional investigation. The engineer determines which direction to pursue and communicates the next task through a prompt. The framework carries the relevant context into that task, including earlier results and the current parameter bounds. Numerical outputs remain associated with the resource that produced them, providing a basis for distinguishing stored simulation data, ML predictions, and newly verified process simulation results.

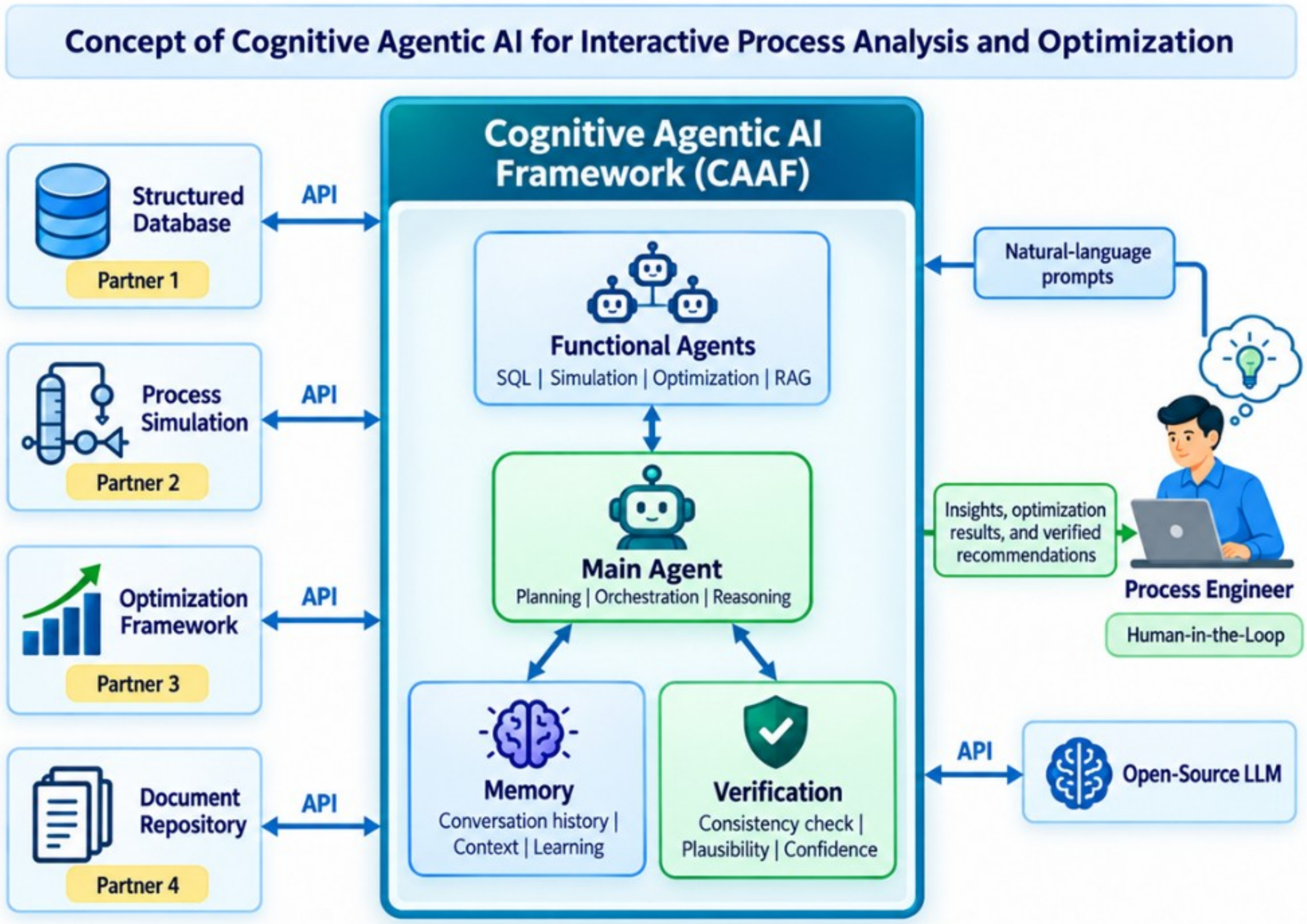

**Figure 1.** Overall concept of CAAF for interactive process analysis and optimization. Four complementary engineering resources connect to the framework through interfaces. Engineers supply domain knowledge and objectives, review returned evidence, and guide subsequent requests. The process topology remains fixed while selected operating and equipment parameters may vary.

## 2.2. Cognitive Agent Architecture

The architecture comprises a Main Agent, functional agents, providers and domain tools, knowledge-synthesis components, verification and memory, and a Knowledge Guidance Layer, as shown in Figure 2. LangGraph provides the orchestration framework for coordinating the agents and their task state [29]. The implementation uses the Gemma 4-31B open-weight language-model family[30]. The Gemma 4-31B Model is accessed through hosted APIs Google AI Studio API [30] rather than local deployment. Harness-engineering principles are implemented through the organization of these components and their interfaces. The term cognitive refers here to the coordinated functions of interpreting engineering requests, planning tasks, selecting resources, synthesizing evidence, and maintaining investigation context. Numerical calculations are performed by the corresponding domain tools, and the agents organize how these calculations address the engineer's question.

### 2.2.1. Main Agent

The Main Agent coordinates the investigation. It interprets the user's objective, identifies the relevant process variables and resources, and determines the analyses needed to address the request. The Knowledge Guidance Layer supplies contextual information such as variable terminology, database structure, and process relationships. This information helps transform a domain-oriented question into tasks that can be assigned to the appropriate functional agents.

Coordination follows an iterative action–observation pattern consistent with ReAct [31,32]. Results returned from one task inform the next task within the current investigation. For example, a database comparison can lead to a request for an explanatory document, or optimization outputs can lead to simulation of selected candidates. Across conversational turns, the Main Agent retains the connection to previous findings so that the engineer can refer to an earlier plot, candidate set, or search region. Revised objectives and bounds are taken from the engineer's subsequent instructions and incorporated into the next computational request.

### 2.2.2. Functional Agents

Four functional agents provide the domain capabilities used in this study: the SQL Agent, Simulation Agent, Optimization Agent, and RAG Summarizer Agent. Each translates a task assigned by the Main Agent into a structured request for its associated provider. Their roles reflect the different information and execution requirements of the connected engineering resources.

The SQL Agent identifies the process records required for analysis, generates SQL queries, and communicates with the SQL provider to retrieve the necessary data. The Simulation Agent coordinates evaluation of specified conditions in the fixed process simulation model and receives the resulting outputs and execution status. The Optimization Agent formulates requests to the existing optimization platform, including the selected objectives and parameter ranges. The RAG Summarizer Agent retrieves and summarizes document evidence relevant to the investigation. These agents allow a single engineering question to draw on several complementary resources while preserving the origin of each returned result. The modular organization also provides a practical basis for adding further resource capabilities as the application develops.

### 2.2.3. Providers and Domain Tools

Providers connect the functional agents to executable engineering resources. The present implementation uses providers for a SQLite database, a document-retrieval pipeline, a Python-based optimization platform, and a process simulation model accessed through Python. Each provider translates the received request into operations supported by its resource and returns the resulting information or execution status.

The SQL provider executes SQL queries, while the document provider retrieves relevant source passages. The optimization provider runs the prepared ML-based optimization workflow. The simulation provider applies the requested input conditions to process simulation, executes the model, and extracts the requested outputs after assessing the returned run status. This division gives the high-level agents a consistent route to specialist resources while keeping the detailed numerical procedures within their existing implementations. Provider outputs supply the evidence used by the analytical, visualization, and synthesis components to answer the original engineering request.

### 2.2.4. Knowledge Synthesis

Knowledge synthesis converts heterogeneous tool outputs into an engineering interpretation. The Data Analyst Agent processes structured and numerical results to produce summaries, comparisons, and relationships among process variables. The Visualization Agent presents the relevant quantities graphically. The Synthesizer Agent combines these outputs with retrieved document evidence and the investigation objective, then returns the integrated response to the Main Agent.

The interpretation retains the context needed to assess each finding, including the analyzed parameter region and the source of reported values. Statistical associations derived from the database describe the sampled simulation results, while literature evidence supports discussion of possible process mechanisms.

Optimization outputs describe the performance predicted by the ML model, and process simulation evaluations provide the corresponding simulator-based assessment for selected conditions. Bringing these forms of evidence together helps the engineer examine trade-offs and identify useful follow-up questions. The next interaction may investigate a narrower region, compare alternative candidates, or request additional numerical evidence.

### 2.2.5. Verification, Failure Handling, and Memory Update

In CAAF, the Verification module focuses on assessing the consistency and reliability of task execution at the workflow level. The verification process evaluates whether the tasks performed and the results returned by the agents are consistent with the engineer's original request, including the specified variables, analysis objectives, and investigation scope. This process is supported by the Knowledge Guidance Layer, which provides domain-specific knowledge, database structures, and relationships among process parameters to ensure consistent interpretation of engineering requests and results. In addition, the Verification module utilizes information stored in Memory, including investigation context, reasoning history, intermediate results, and previous interactions, to assess consistency and continuity across successive tasks. By combining the Knowledge Guidance Layer and Memory, the framework can identify results that are inconsistent with the investigation objectives or previous findings, providing relevant information for engineers to review and refine subsequent requests.

CAAF also incorporates a failure-handling mechanism to monitor the execution status of process simulations. If a simulation fails to converge or encounters an execution error, the framework reports the unsuccessful condition and marks the corresponding results as invalid. Numerical outputs from failed simulations, including any values retained from a previously converged run, are excluded from the valid result set. The workflow then automatically proceeds to the next simulation condition without interrupting the overall evaluation process. The execution status of both successful and failed simulations is recorded and reported, allowing engineers to identify problematic conditions and determine whether further analysis or adjustments are required.

In addition, CAAF incorporates a logging mechanism that records and provides feedback throughout its operation, including workflow-level processing and reasoning steps, planning, agent orchestration, tool execution, generated code, and outputs from individual agents. This information is systematically recorded, allowing users to monitor task execution in real time or review previous processing steps, executed commands, and intermediate results when necessary. This mechanism enhances the transparency and traceability of the framework, enabling engineers to better understand how individual agents collaborate to execute tasks and generate results. Furthermore, the recorded execution history and feedback allow users to identify potential inconsistencies, execution errors, or unsupported outputs (hallucinations) produced by individual agents. Based on this information, engineers can investigate potential causes and refine prompts or execution conditions to improve the reliability of subsequent interactions. This approach facilitates human oversight and provides a foundation for the continuous evaluation, monitoring, and improvement of the Cognitive Agentic AI framework.

### 2.2.6. Knowledge Guidance Layer

The Knowledge Guidance Layer supplies the domain context used throughout request interpretation, task formulation, and result assessment. Its resources include database schemas, process terminology, descriptions of units and streams, parameter relationships, and relevant engineering knowledge. These resources connect the language used by the engineer with the variables and capabilities exposed by the providers.

For example, a question about feed composition, conversion, pressure, or energy cost must be associated with the appropriate quantities before a query or computational request is formed. The same contextual information helps the Main Agent select resources and helps the Synthesizer Agent explain the resulting relationships. Guidance thus provides a shared interpretation of the process across successive tasks. Specified numerical bounds are carried into the relevant computational request, while domain descriptions support interpretation of the resulting evidence. In combination with memory and verification, this layer maintains a coherent connection between engineering intent and the calculations performed during the investigation.

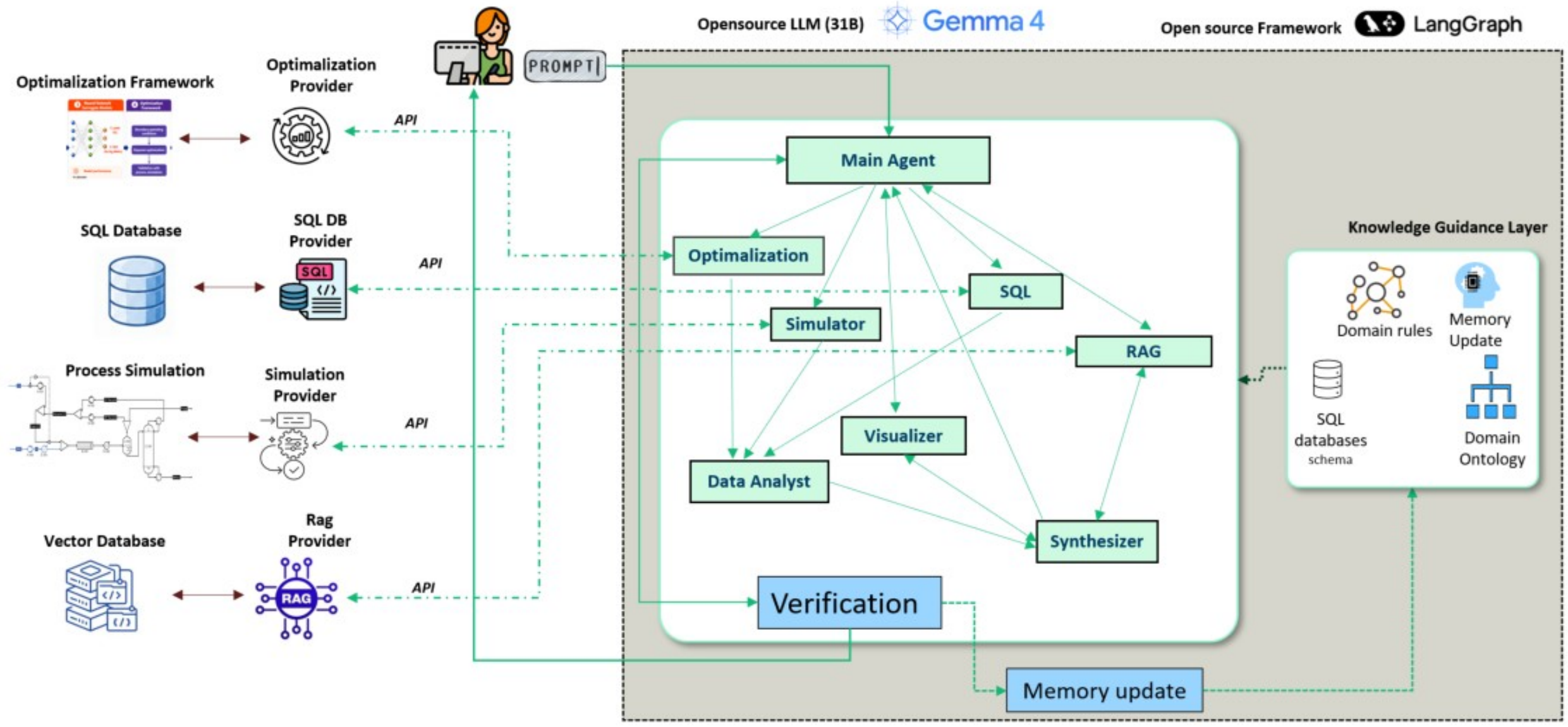

**Figure 2.** Cognitive agent architecture is based on harness-engineering principles for coordinating interactive process investigations. Functional agents interact with specialized Providers and domain-specific resources to support data analysis, simulation, optimization, and knowledge synthesis. The Verification and Memory modules ensure workflow consistency and maintain investigation context across successive interactions.

## 3. Case Study

The case study applies CAAF to interactive analysis and optimization of a methyl acetate (MeAce) production process. It connects a MeAce process simulation flowsheet, two structured simulation datasets, an existing machine-learning-based optimization resource, and a scientific-document repository. The investigation progresses from interpreting process relationships to exploring objective trade-offs, refining the search domain, and checking selected candidates in the process simulation flowsheet. This sequence provides a common engineering context in which successive user requests draw on different computational resources while retaining the results of preceding analyses.

### 3.1. Methyl Acetate Production Process

The process produces methyl acetate through the carbonylation of dimethyl ether (DME) with carbon monoxide (CO). The configuration follows the carbonylation section described by Diemer and Luyben and represented in the process simulation flowsheet collection [1,33]  It includes feed compression, catalytic reaction, cooling, phase separation, distillation, and gas recycle. Fresh and recycled reactants enter the reactor, after which cooling and phase separation provide a liquid stream for purification. The gas recycles and purge branches govern reactant recovery and the removal of non-reacting material. The flowsheet therefore connects reaction performance with downstream separation and compression requirements. Five inputs are investigated: reactor temperature, reactor length, pressure, CO molar flow, and splitter factor. The input vector of key parameters is consequently described as a set of operating and equipment parameters. Throughout database generation, optimization, and candidate re-evaluation, the unit-operation arrangement and stream connections remain fixed. Parameter changes are applied to this common flowsheet.

Two existing process-performance indicators define the optimization objectives. Carbon conversion yield, (C-Yield) is reported as a dimensionless fraction, while specific energy consumption (C-SEC) denotes the cost of specific energy consumption per kilogram of methyl acetate, expressed in EUR/kg. Their calculation is supplied by the numerical resources using the metric definitions adopted for the case study. Additional outputs, including methyl acetate mass flow, reboiler duty, and compressor duties, support interpretation of the candidate conditions.

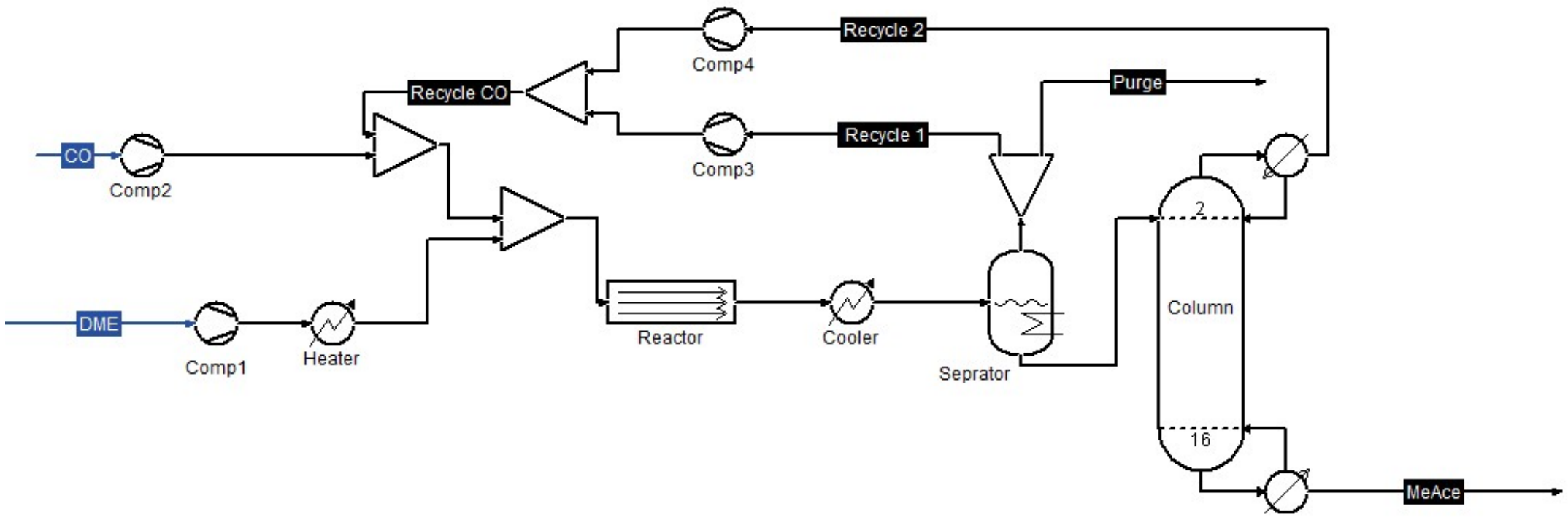

**Figure 3.** Process simulation flowsheet for the methyl acetate carbonylation case, based on the literature-derived sample model [1,33]

## 3.2. Federated Knowledge Base

The knowledge environment is organized into four partner roles, each responsible for a distinct engineering resource: process simulation, structured data, optimization, and document retrieval. These roles define the resource boundaries used in the demonstration. Dedicated Providers expose their capabilities to CAAF through Python-based interfaces. The organization allows the investigation to combine numerical outputs and documentary evidence while preserving the origin and function of each contribution. The simulation model supplies the common modelling basis for the structured datasets and optimization models.

### 3.2.1. Partner 1 – Process Simulation Models

Partner 1 provides the steady-state methyl acetate model implemented in a steady-state process simulations CAPE-OPEN to CAPE-OPEN (COCO). COCO is an open-source, CAPE-OPEN–compliant simulation platform that enables integration of thermodynamic packages, unit operation models, and flowsheet solvers within a modular and extensible framework. The model adopts the process configuration, reaction scheme, and kinetic basis of the literature-derived example [1,33]. A Win32 interface connects COCO to Python, enabling the simulation resource to receive input conditions, modify the corresponding model parameters, execute a calculation, and retrieve the requested outputs. This interface makes the same numerical model accessible during resource preparation and subsequent interactive investigations.

COCO serves two complementary purposes. During preparation, it generates the process cases used to establish the structured databases and train the process surrogates. During interactive use, it evaluates selected conditions proposed by the optimization resource. Each candidate is transferred with its complete input vector so that the surrogate prediction and simulator result refer to the same requested conditions.

Direct re-evaluation also returns process quantities beyond the surrogate targets, allowing the engineer to examine product flow and equipment duties when interpreting the optimization outcome.

The simulation workflow assesses convergence before accepting candidate outputs. Under the reporting-and-continuation policy described in Section 2.2.5, a non-converged condition is reported as unverified, its numerical outputs are excluded from accepted results, and subsequent requested conditions are evaluated. This procedure separates unsuccessful attempts from usable results. Determining the cause of non-convergence requires further examination of the model and solution procedure.

### 3.2.2. Partner 2 – Structured Databases

Partner 2 provides a SQLite database assembled from approximately 6636 COCO simulation cases. Across its two datasets, the resource contains more than 216 columns describing process inputs, stream properties, unit-operation outputs, and performance indicators. The records are simulation-generated engineering data. The use of an established process model provides the basis for calculating these records even though experimental pilot-plant measurements are not included in the present case study.

Dataset 1 contains approximately 3906 scenario-based cases. The prescribed scenarios represent variations in operating conditions and changes intended to examine process fluctuations, reduced catalyst performance, and troubleshooting-related questions. They provide a resource for comparing process responses across multiple units, such as changes in reactor conditions alongside distillation temperatures, heat duties, and product flow. Because the underlying calculations are steady-state, the ordered records represent a sequence of scenario evaluations. Their interpretation concerns differences between computed states.

Dataset 2 contains approximately 2730 design-of-experiments (DOE) cases generated by varying the five selected inputs. It supports analysis of multivariable relationships and development of the ML models used by the optimization resource. These activities share the simulation data but address different questions: exploratory analysis characterizes the sampled process responses, whereas surrogate development establishes a predictive input-output representation for subsequent searches. The SQL interface retrieves the records and variables required for each task.

The adopted data-quality rule admits only completed simulations with usable target outputs into surrogate development. Keeping the scenario and DOE datasets identifiable also allows the framework to select a dataset according to the engineering request: scenario comparisons use Dataset 1, while the principal parameter analysis and surrogate preparation use Dataset 2.

### 3.2.3. Partner 3 – Optimization Framework

Partner 3 provides an in-house Python-based Bayesian Optimization (BO) platform supporting both single- and multi-objective optimization as described in detail in our work [34]. The platform integrates ML surrogate models trained and validated using the DOE dataset from Partner 2 to predict carbon conversion yield (C-Yield) and the cost of specific energy consumption (C-SEC). The optimization objectives include maximizing C-Yield individually or simultaneously maximizing C-Yield and minimizing C-SEC.

Through CAAF, engineers can specify optimization objectives, parameter bounds, and the number of optimization iterations using natural-language prompts. The platform performs numerical optimization and returns candidate conditions with their predicted performance. Selected candidates are subsequently re-evaluated using the COCO process simulation model to assess the agreement between ML predictions and simulation results.

### 3.2.4. Partner 4 – Retrieval-Augmented Generation (RAG)

Partner 4 provides an unstructured repository containing approximately 121 pages of PDF documents, including scientific publications, technical reports, and operating documentation. The repository supplies contextual information on methyl acetate production, carbonylation, catalyst behaviour, separation, recycle, and process performance. Its role is to support the interpretation of numerical findings and help the engineer connect observed relationships with relevant process knowledge.

The retrieval pipeline is implemented using LangChain [29] and combines semantic retrieval, parent-child chunking, embedding models, reranking, and reconstruction of document context. Processed document content is stored in a vector database. In response to a query, the RAG provider retrieves relevant passages and returns the associated source information for synthesis with outputs from the other resources. The resulting explanation can therefore connect a computed trend with an identifiable literature basis. Retrieved mechanisms are interpreted in relation to the assumptions of the implemented process model, and their relevance can be examined through subsequent analysis or simulation requests.

### 3.2.5. Providers

Each functional agent communicates with its resource through a dedicated Provider. The provider receives the requested task, checks and transforms its inputs into executable operations, and returns the resulting data and execution status. Python supplies the common integration environment, while each resource retains its own numerical or retrieval procedure. This division connects the engineer's natural-language objective with explicit operations on the case-study resources.

The SQL provider translates data requests into database queries and formats the retrieved records or summaries. The Simulation provider applies candidate inputs to the fixed COCO flowsheet, executes the calculation, and extracts the requested variables. The Optimization provider passes objectives, parameter bounds, and the iteration budget to the existing search platform, then returns candidate conditions and predicted performance. The RAG provider retrieves, ranks, and reconstructs relevant document passages together with their source information.

The outputs preserve their computational roles throughout synthesis: database values describe stored simulation cases, optimization values are ML predictions, and verification values are obtained from direct COCO calculations. The simulation outcome accompanies the relevant candidate in the response returned to CAAF. This organization allows the framework to connect analyses without losing the distinction between a predicted candidate and a successfully re-evaluated one. It also makes the inputs and outputs of each resource available for checking whether the executed task corresponds to the engineer's request.

## 3.3. Prompt Support for Case Study

The case study demonstrates an interactive engineering investigation through 9 sequential prompts, as summarized in Table 1. The detailed prompts and their corresponding results are presented in Supplement Information 1. The investigation begins with analysis of the DOE dataset and scientific literature to identify relationships between process inputs and performance indicators. Subsequent prompts request multivariable visualizations and comparisons of distillation performance across different operating scenarios. The engineer then initiates multi-objective Bayesian optimization over the full parameter domain, using 4,000 optimization iterations to maximize C-Yield and minimize C-SEC.

**Table 1.** Engineering prompt sequence and the evidence used to guide subsequent interaction. The prompts are concise summaries of the case-study requests.

| Prompt set | Prompt summary | Returned evidence or human decision |
|---|---|---|
| 1 | Analyze the influence of operating parameters on C-YIELD and C-SEC using scientific literature and the SQL DOE dataset. | Retrieved literature and database relationships provide the basis for the initial process investigation. |
| 2 | Visualize multivariable relationships between operating parameters and C-YIELD using the DOE dataset. | Parameter–yield distributions reveal operating trends and interactions, providing initial process insights. |
| 3 | Analyze long-term pilot test data (14 March–15 May 2026) by comparing methyl acetate production and operating parameters over time. | Time-series visualizations support the assessment of process stability and production performance under nominally constant operating conditions. |
| 4 | Maximize C-YIELD and minimize C-SEC using 4000 Bayesian optimization iterations across five design and operating variables. | Predicted objective values and Pareto-optimal candidates establish the initial optimization results within the full search domain. |
| 5 | Visualize the relationships between operating parameters, C-YIELD, and C-SEC using color-coded scatter plots. | Multivariable distributions reveal favorable operating regions and trade-offs between carbon yield and energy cost. |
| 6 | Repeat the multi-objective optimization with 4,000 trials under restricted pressure, reactor length, and splitter factor limits. | Engineering judgment refines the search domain to explore potential CAPEX and OPEX reductions while maintaining favorable process performance. |
| 7 | Investigate parameter–performance relationships within the selected high-yield, low-energy-cost region. | Focused visualizations provide deeper insights into operating-variable interactions and objective trade-offs within the restricted region. |
| 8 | Verify Pareto-front candidates using COCO simulation and compare predicted C-YIELD values through a parity plot. | COCO simulation results and parity analysis assess prediction accuracy and distinguish verified solutions from unverified candidates. |
| 9 | Evaluate MeAce production and equipment energy requirements for COCO-verified Pareto-front solutions. | Equipment-level performance plots and a summary table provide additional evidence for engineering evaluation and operating-condition selection. |

Based on the initial optimization results, the engineer refines the operating ranges and requests a second multi-objective optimization within the restricted domain. CAAF translates these instructions into computational tasks while maintaining the context of previous analyses. Finally, selected optimization candidates are re-evaluated using the COCO process simulator to compare ML predictions with simulation results and assess convergence status. This sequence demonstrates how natural-language prompts and human-guided decisions support continuous process analysis, optimization, and simulation-based verification within a unified framework.

## 4. Results and Discussion

### 4.1. Prompt-Driven Interactive Process Analysis and Optimization

The case study demonstrates a continuous, prompt-driven engineering investigation rather than a sequence of isolated numerical tasks. As summarized in Table 1, six successive prompts connect structured-data analysis, literature retrieval, multivariable visualization, multi-objective optimization, investigation of the Pareto-optimal region, and COCO-based verification. Through these interactions, the methyl acetate flowsheet remains fixed; the engineer varies five operating and equipment parameters and uses intermediate evidence to determine the next investigation step. CAAF coordinates the specialized resources and preserves the context of previous findings, while numerical predictions and calculations remain the responsibility of the corresponding engineering tools. A more comprehensive evaluation of CAAF's reliability and agent-level performance has been reported in Supplement Information 2 and discussed in section 4.2.

### 4.2.1 Interactive Multivariable and Scenario-Based Process Analysis

To investigate the relationships between operating conditions and process performance, the user interacted with CAAF through Prompts 1 and 2 (see table 1). These prompts were applied to the DOE dataset (Dataset 2) to explore the relationships between five operating parameters and two process objectives: C-Yield and C-Sec. In response, CAAF retrieved and analyzed the relevant simulation data and generated a correlation heatmap (Figure 4) and multivariable visualizations (Figure 5).

Figure 4 presents the Pearson correlation coefficients between operating parameters and process objectives. Reactor temperature exhibits the strongest positive correlation with carbon efficiency ($r = 0.59$), followed by CO molar flow ($r = 0.40$) and reactor length ($r = 0.24$). Conversely, the splitter factor shows a negative correlation ($r = -0.28$), while compressor pressure has a weak linear association ($r = -0.07$). For methyl acetate-specific energy consumption, reactor temperature and CO molar flow show negative correlations of −0.38 and −0.39, respectively, whereas compressor pressure exhibits a positive correlation of 0.32. These results provide an initial indication of the relative importance and direction of parameter associations within the investigated design space, without establishing causal relationships.

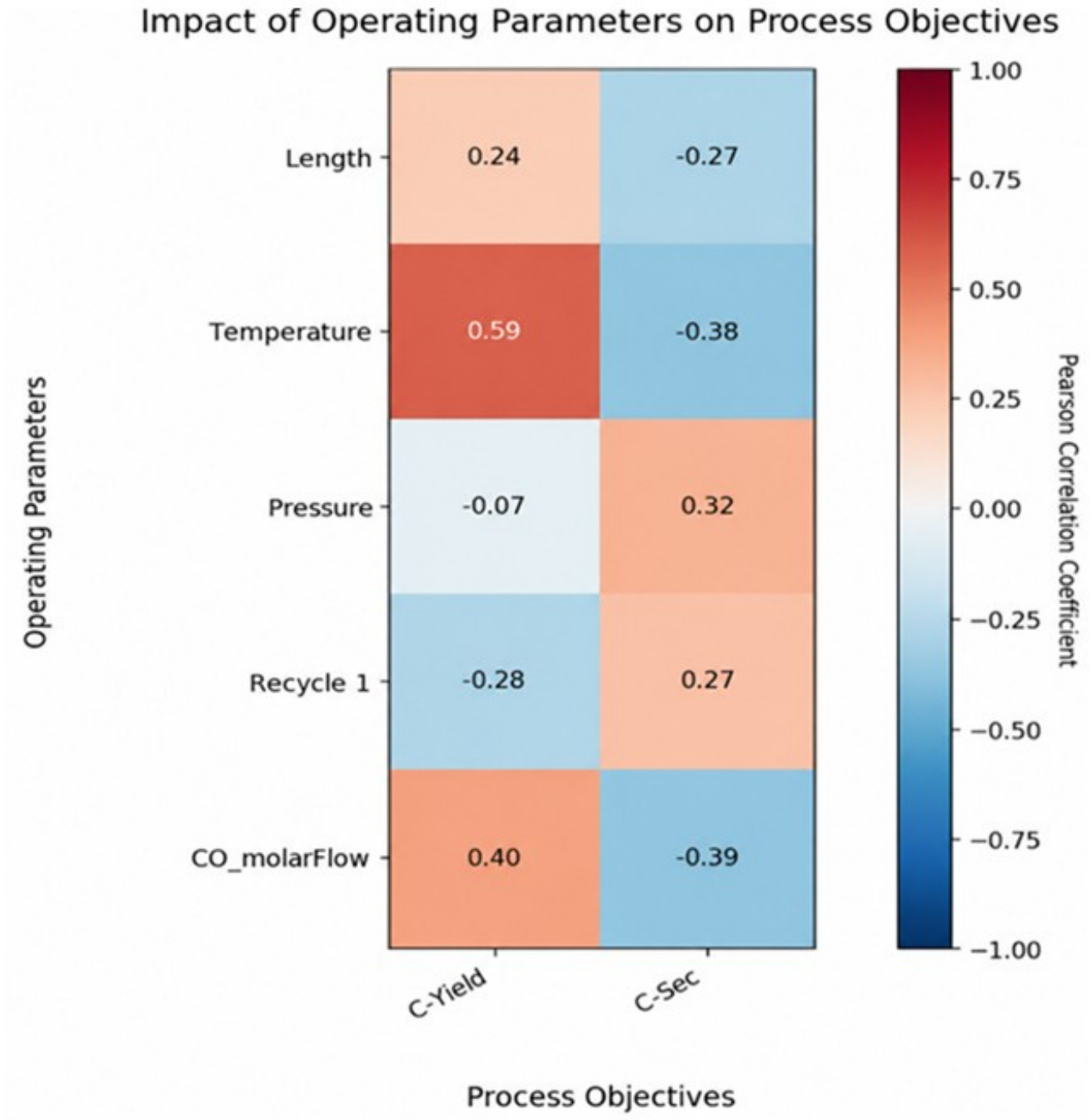

Figure 4. Pearson correlations between operating parameters and process objectives generated by CAAF.

To further investigate these relationships, CAAF generated multivariable scatter plots showing the distribution of C-YIELD across combinations of reactor length, temperature, pressure, CO molar flow, and splitter factor (Figure 5). The results reveal a pronounced dependence of the yield distribution on reactor temperature, with higher yield values occurring more frequently at elevated temperatures. Reactor length also influences the distribution of attainable yields, while the variation associated with CO molar flow, pressure, and splitter factor illustrates the multivariable nature of process performance. The substantial differences in yield observed at similar values of individual operating parameters highlight the importance of examining parameter combinations rather than relying solely on one-variable relationships. These visualizations allow users to identify potentially favorable operating regions for subsequent detailed analysis and optimization.

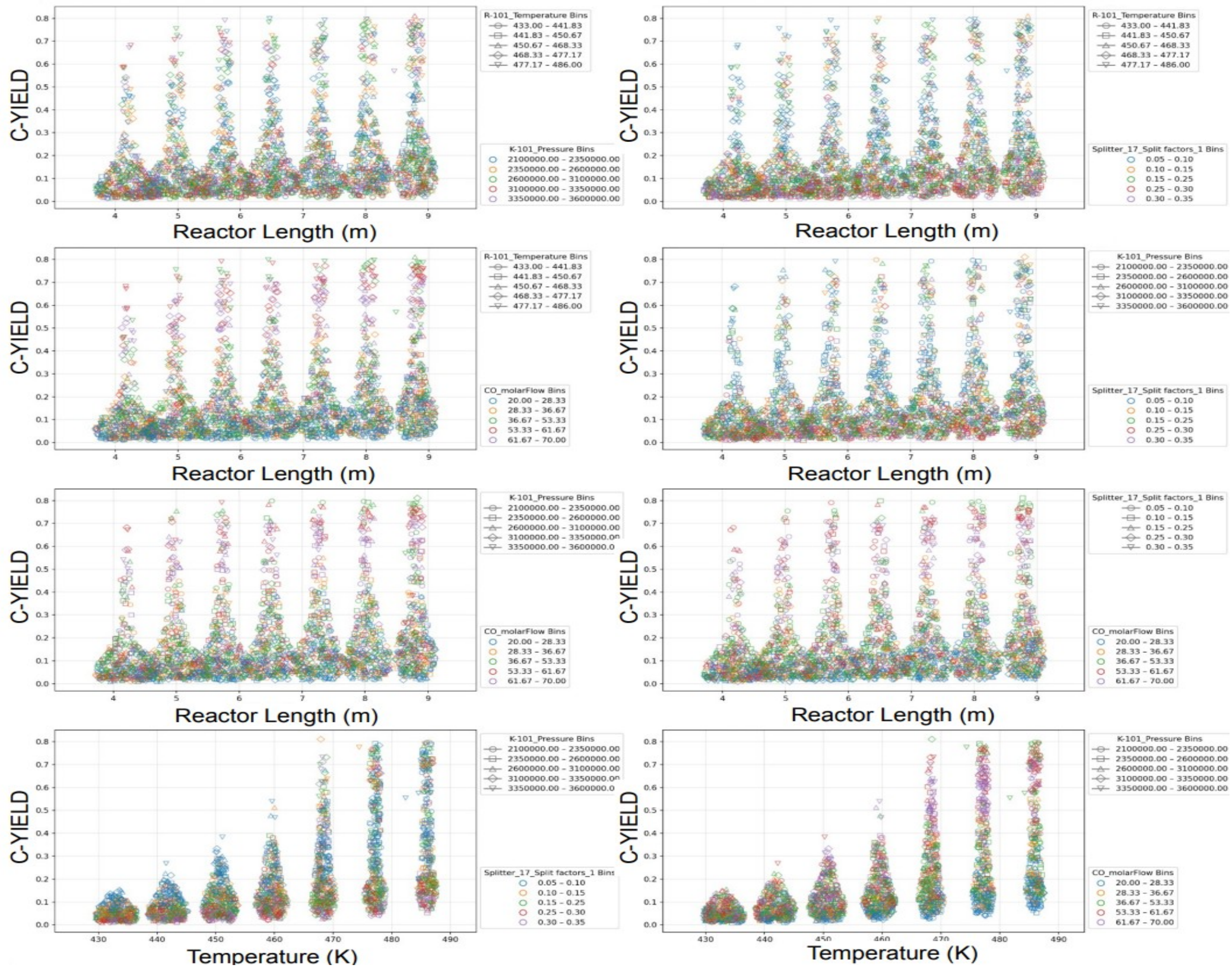

Figure 5. Multivariable relationships between operating parameters and carbon conversion yield (C-YIELD), generated by CAAF from the DOE simulation dataset.

To further interpret the observed relationships, CAAF retrieved relevant scientific knowledge from the RAG literature repository to provide engineering explanations for the DOE-based findings. The positive correlations of reactor temperature ($r = 0.593$) and CO molar flow ($r = 0.397$) with C-YIELD were consistent with literature-based explanations involving temperature-dependent reaction kinetics and increased CO availability for carbonylation. Their negative correlations with C-SEC ($r = -0.381$ and $-0.387$, respectively) suggest that improved conversion and production efficiency may reduce energy consumption per unit of methyl acetate produced. Conversely, the positive correlation between compressor pressure and C-SEC ($r = 0.318$) is consistent with increased compression energy requirements. The multivariable analysis further revealed favorable C-YIELD at higher reactor temperatures and CO flow rates. These findings illustrate how CAAF integrates quantitative DOE analysis with literature-derived engineering knowledge to support process understanding and the interpretation of operating-parameter effects.

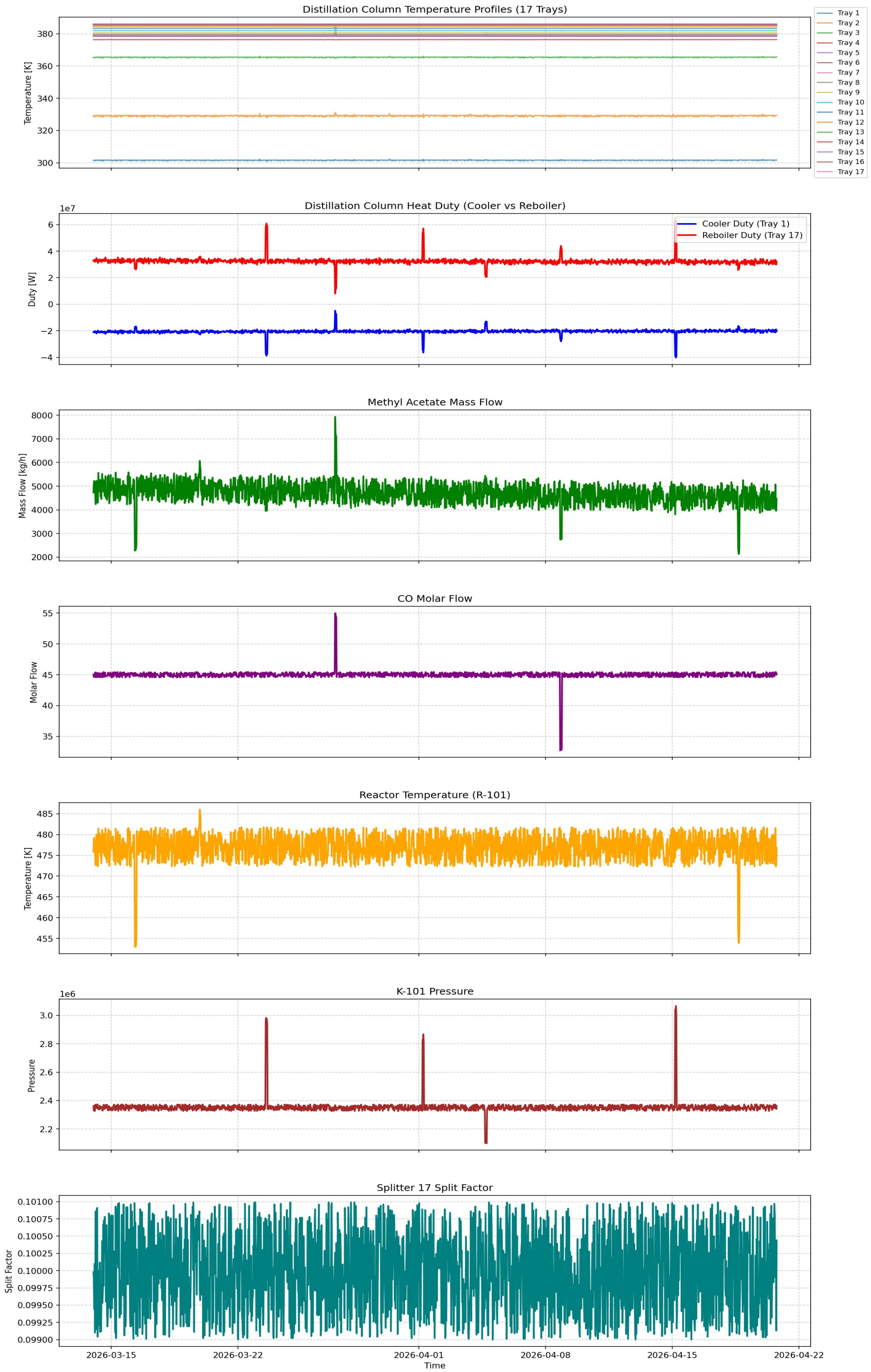

Figure 6. C-101 distillation-column tray temperatures, cooler and reboiler heat duties, and methyl acetate product mass flow across steady-state simulation scenarios (see figure 3), visualized by CAAF.

In a subsequent prompt (prompt 3 in table 1), the user requested another analysis from the reactor operating conditions to the downstream distillation section. In response, CAAF retrieved and visualized time-series simulation data from the operating dataset (Dataset 1) for the C-101 distillation column, including temperature profiles at 17 tray locations, cooler and reboiler heat duties, and methyl acetate product mass flow rate over time (Figure 6). The results show that the temperature at each tray location remains relatively consistent across the investigated scenarios, preserving the overall temperature distribution within the column. In contrast, the heat duties exhibit scenario-dependent variations, including several pronounced deviations, while methyl acetate production also varies considerably. These observations indicate that similar column temperature profiles can be associated with different energy requirements and product flow rates.

In a subsequent prompt, the user requested CAAF to further investigate the influence of upstream reactor operating conditions on downstream distillation performance. By retaining and reusing the analytical context and results from previous interactions, CAAF extended the earlier investigation to identify relationships between reactor conditions, methyl acetate production, and equipment energy requirements. The analysis revealed a strong positive correlation between reactor temperature (R-101) and methyl acetate production rate ($r = 0.877$), while compressor pressure (K-101) was strongly correlated with reboiler heat duty ($r = 0.843$). Time-series analysis further indicated that fluctuations in reactor temperature and splitter ratio coincided with variations in distillation performance. Although these correlations do not establish direct causality, they suggest important relationships between upstream operating conditions and downstream process behavior. This demonstrates CAAF's ability to maintain analytical continuity across successive prompts, reuse previously generated evidence, and progressively develop deeper process insights without requiring users to repeat earlier analyses.

The demonstrated data analysis and literature retrieval capabilities highlight CAAF's ability to integrate and interact with multiple information sources, including structured databases and vector databases, while leveraging LLM-based reasoning to support engineering interpretation. By combining quantitative data analysis with scientific knowledge retrieved from the literature, CAAF enables users to rapidly gain insights into complex interactions among process parameters and investigate how abnormal operating conditions may affect overall process performance. This integrated approach demonstrates the potential of CAAF to provide timely, evidence-based explanations, facilitating a more efficient and informed understanding of process behavior and operational disturbances.

### 4.2.3. Prompt-Driven Multi-Objective Optimization and Human-Guided Refinement

Optimization operating parameters for process design is a critical step in improving process performance and economic feasibility. A key objective is to identify high-performing configurations that maximize product yield while minimizing energy-related costs, with capital expenditure (CAPEX) and operating expenditure (OPEX) remaining important considerations for the overall techno-economic assessment (TEA). However, process performance is typically governed by multiple interdependent design and operating variables with complex nonlinear interactions. Exploring these relationships and identifying favorable combinations across a multidimensional search space can require a substantial number of computational evaluations. Consequently, efficient optimization strategies are essential for systematically exploring the operating domain, identifying promising solutions, and quantifying trade-offs between competing performance objectives.

Following the exploratory process analyses, the user requested CAAF to perform ML-based multi-objective optimization to maximize C-YIELD while minimizing C-SEC, with priority given to improving C-YIELD. The initial optimization was conducted over a relatively broad search domain involving five design and operating variables: reactor length (4.2–8.8 m), reactor temperature (433.0–486.0 K), CO molar flow rate (30–70 mol $s^{-1}$), compressor pressure (2.1–3.6×$10^6$ Pa), and splitter factor (0.05–0.35). A computational budget of 4,000 optimization iterations was specified to systematically explore the multidimensional parameter space and identify favorable combinations of the two objectives. CAAF interpreted the user-defined optimization objectives, parameter bounds, and computational budget, translated them into the required tool inputs, and invoked the existing Bayesian optimization platform to explore candidate process conditions. The resulting optimization data were subsequently retrieved, organized, and visualized by CAAF to facilitate engineering interpretation and further investigation (Figure 7).

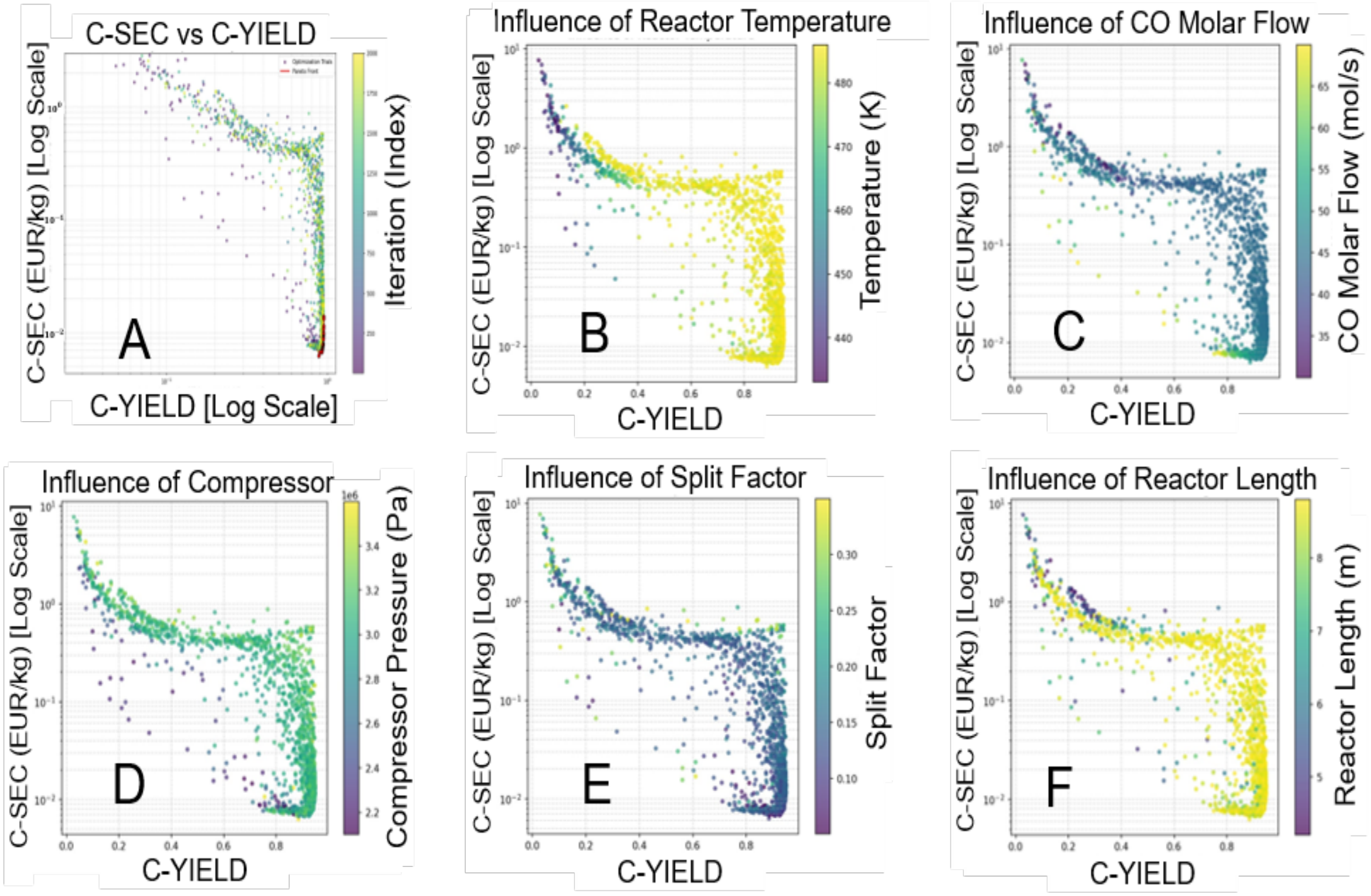

Figure 7. Full-domain ML-based multi-objective optimization generated by CAAF. (A) Evaluated objective pairs and highlighted non-dominated solutions. (B–F) Distributions of reactor temperature, CO molar flow, compressor pressure, splitter factor, and reactor length, respectively. C-SEC is expressed in EUR/kg of methyl acetate on a logarithmic axis.

Figure 7A presents the distribution of evaluated objective pairs and the identified non-dominated solutions, with the maximum predicted C-YIELD approaching 0.95 and C-SEC values reaching approximately 6.3 × $10^{-3}$ EUR $kg^{-1}$ of MeAce. The overall distribution demonstrates that certain combinations of design and operating conditions can simultaneously improve carbon conversion yield and reduce specific energy cost relative to less favorable alternatives. The non-dominated solutions define the Pareto front, representing the best trade-offs identified between the competing objectives within the evaluated search space. Figures 7B–F further illustrate the relationships between the five optimization variables and the objective distributions. Higher reactor temperatures and longer reactor lengths are generally associated with increased C-YIELD, while lower splitter factors frequently characterize favorable solutions. In contrast, compressor pressure and CO molar flow exhibit more complex relationships with process performance, suggesting that their effects depend on interactions with the other design and operating variables. These distributions provide valuable insights into the relationships between process conditions and performance, supporting the identification of promising operating regions for subsequent engineering evaluation.

Importantly, the Pareto front highlighted in red in Figure 7A reveals the fundamental trade-off between C-YIELD and C-SEC. Along this frontier, further improvements in carbon conversion yield may require an increase in specific energy cost, indicating that maximizing yield does not necessarily correspond to the most economically favorable operating condition. This trade-off information is particularly valuable for engineering decision-making, as the preferred solution depends on the relative importance assigned to product yield, energy cost, and broader economic considerations. Rather than providing only a single optimized solution, CAAF enables the user to examine alternative operating conditions, interpret the associated performance compromises, and make informed decisions based on engineering priorities. This illustrates the value of integrating ML-based multi-objective optimization with interactive, agent-orchestrated analysis to support process understanding and decision-making.

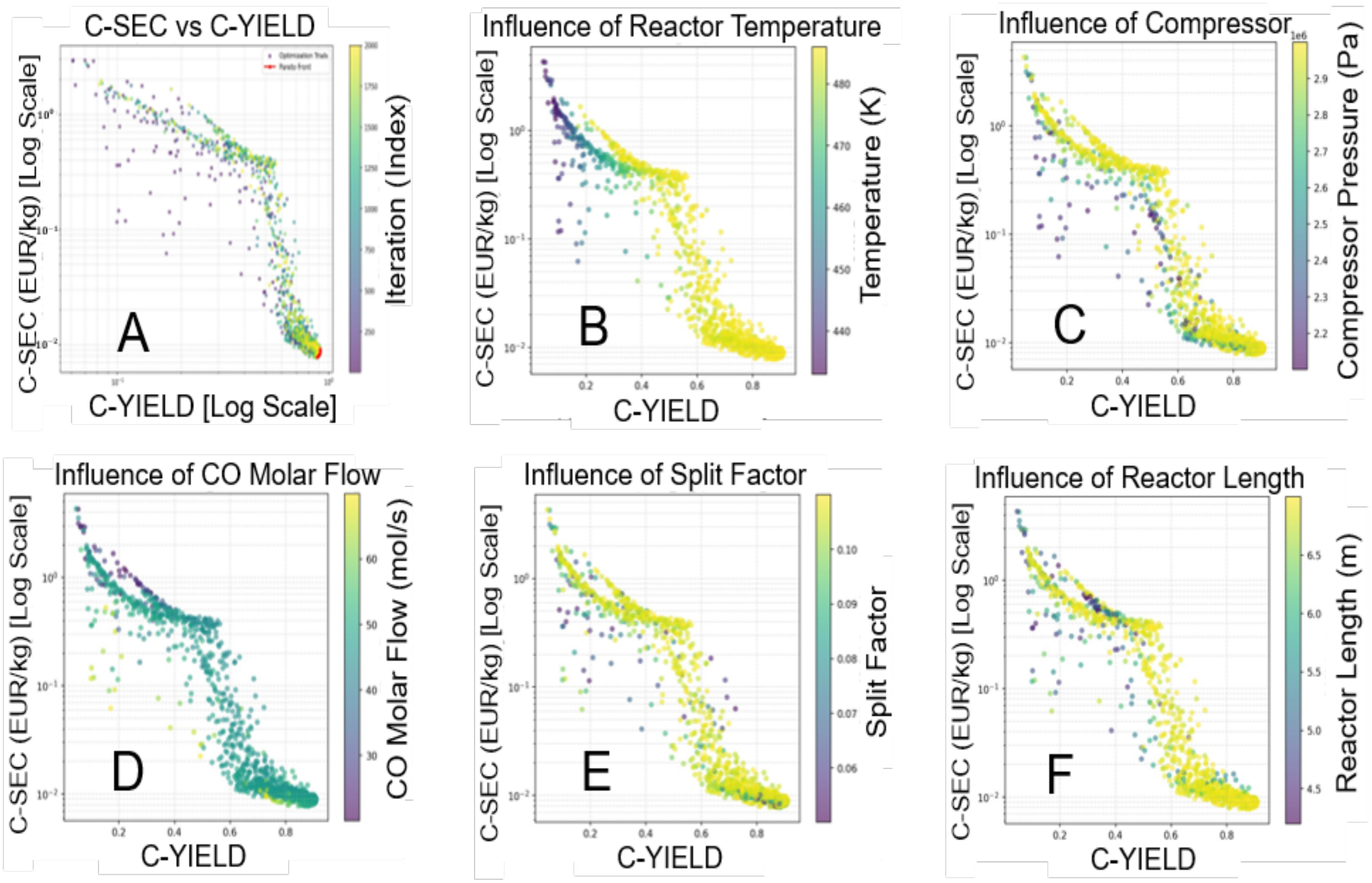

Figure 8. ML-based multi-objective optimization within the human-specified restricted operating region, generated by CAAF. (A) Evaluated objective pairs and highlighted non-dominated solutions. (B–F)

Following the initial optimization, the user initiated a second optimization scenario to explore opportunities for reducing capital and operating expenditures (CAPEX and OPEX) while maintaining the original objectives of maximizing C-YIELD and minimizing C-SEC. Based on engineering judgment and insights obtained from the first optimization, the user introduced additional constraints through a subsequent natural-language prompt, limiting compressor pressure to 3 MPa, reactor length to 7 m, and splitter factor to 0.11. These restrictions were intended to investigate whether favorable process performance could still be achieved within a more constrained design and operating domain, potentially reducing equipment investment and operating costs. This intermediate human intervention demonstrates the interactive capabilities of CAAF, allowing engineers to refine optimization scenarios based on their technical knowledge and economic priorities without modifying the underlying optimization objectives.

CAAF interpreted the updated requirements, translated them into the corresponding optimization constraints, and automatically executed a second Bayesian optimization search. The resulting objective distributions and relationships with the operating variables are presented in Figure 8.

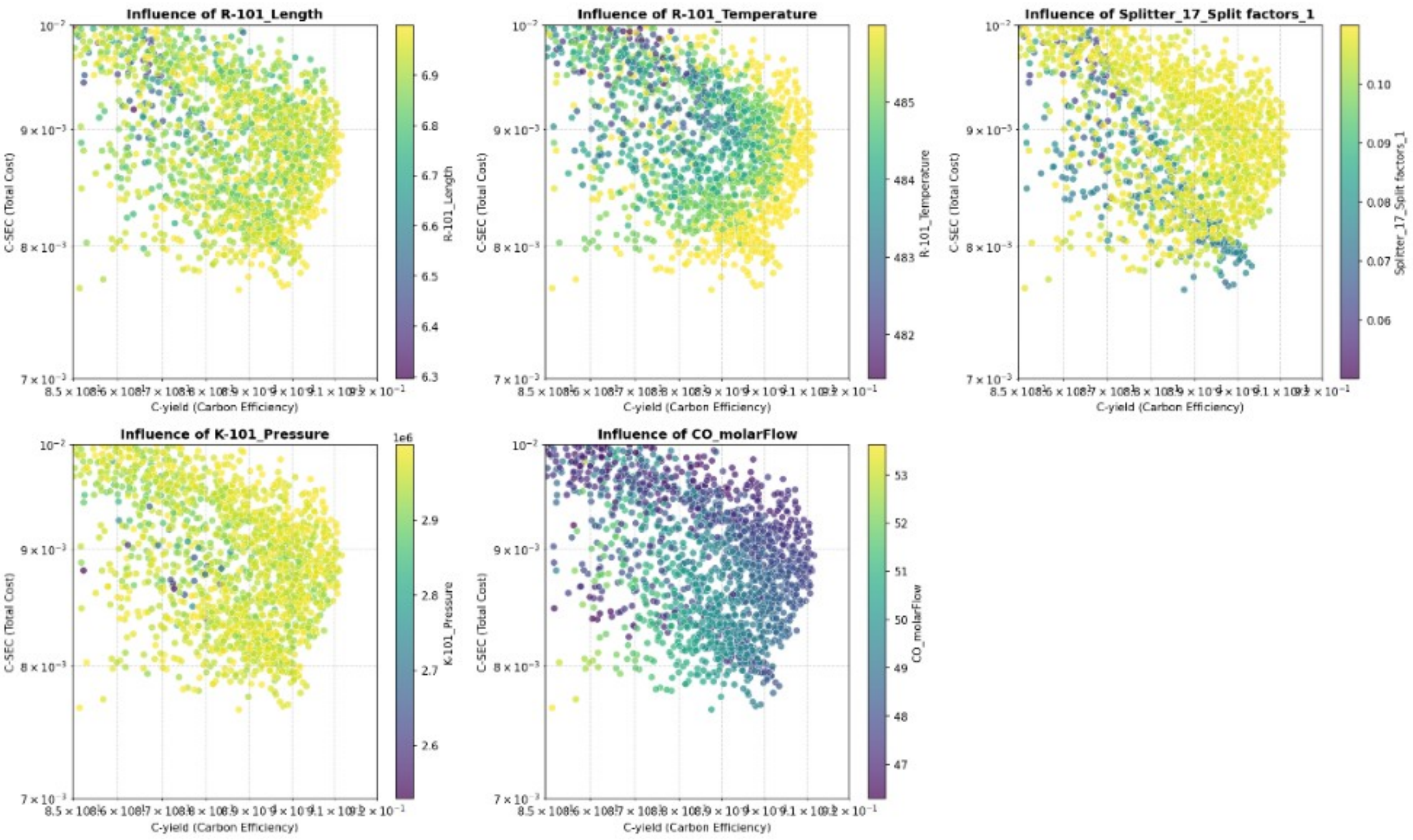

Figure 9. Distribution of operating parameters within the narrow optimization region, generated by CAAF. The relationships between C-YIELD and C-SEC are visualized using five operating parameters: reactor length, reactor temperature, splitter factor, compressor pressure, and CO molar flow.

The constrained optimization identified promising combinations of design and operating conditions, achieving a maximum predicted C-YIELD of approximately 0.91 while maintaining relatively low C-SEC values of $7.7\times 10^{-3}$ EUR $kg^{-1}$ of MeAce (Figure 8A). Although the maximum predicted yield was slightly lower than that obtained from the initial optimization (0.95), the results indicate that relatively high process performance can still be maintained within the more restrictive operating domain. Figures 8B–F illustrate the corresponding distributions of reactor temperature, compressor pressure, CO molar flow rate, splitter factor, and reactor length, respectively. Elevated reactor temperatures remained associated with favorable objective values, whereas the imposed limitations on reactor length, compressor pressure, and splitter factor altered the distribution of candidate solutions and narrowed the feasible search space. The resulting solutions therefore provide a more focused set of operating conditions aligned with the user's engineering and economic considerations. Nevertheless, because the feasible domains differ between the two optimization scenarios, their results should not be interpreted as demonstrating equivalent optimal performance. Furthermore, the potential reductions in CAPEX and OPEX are inferred from the imposed engineering constraints rather than directly quantified through a detailed techno-economic assessment.

These results highlight the flexibility and practical value of CAAF in supporting interactive process analysis and optimization. Rather than relying on a single optimization run with predefined constraints, engineers can iteratively explore alternative design and operating scenarios, incorporate engineering judgment, and examine the resulting trade-offs between process performance and potential economic benefits. Importantly, CAAF facilitates this process through natural-language interaction, automatically coordinating the necessary optimization tools and presenting the results for engineering interpretation. In the demonstrated case, the revised optimization scenario was completed and its results made available for analysis within approximately 6 minutes, illustrating the potential of the framework to accelerate scenario evaluation and support timely engineering decision-making. This interactive, human-guided workflow represents an important capability of CAAF, enabling users to progressively refine their optimization strategies and explore practical alternatives without manually reconfiguring the underlying computational workflow.

To investigate the refined operating region in greater detail, CAAF generated additional multivariable visualizations of the relationships between C-YIELD, C-SEC, and the five operating parameters (Figure 9). The results provide a closer examination of candidate solutions within a narrower parameter space, revealing differences in energy cost and operating-variable combinations among solutions with relatively similar carbon yields. Reactor temperature and reactor length exhibit recognizable distributions toward the higher-yield region, while compressor pressure, splitter factor, and CO molar flow show variations across the objective space. These visualizations help identify the operating-variable combinations associated with potentially favorable process performance and support further examination of candidate solutions before simulation-based verification.

Together, these optimization investigations demonstrate the capability of CAAF to support an interactive, human-guided engineering workflow. The framework enables users to initiate multi-objective optimization through natural-language prompts, examine predicted process performance, refine operating constraints based on engineering judgment, and investigate selected regions in greater detail. This iterative interaction maintains continuity between exploratory analysis and optimization while allowing the engineer to retain control over the admissible operating domain. The resulting solutions provide candidates for subsequent verification using the COCO process simulator, where their predicted performance and feasibility can be assessed against the underlying process model.

### 4.2.4. COCO-Based Verification and Engineering Interpretation of Optimization Candidates

Following the human-guided optimization, the user requested CAAF to verify selected Pareto-optimal candidates using the rigorous COCO process simulator. In response to Prompt 6 (table 2), CAAF transferred the corresponding operating parameters, including reactor length, temperature, pressure, CO molar flow, and splitter factor, from the optimization results to the established COCO flowsheet. The simulation interface executed the calculations, checked their execution status, and retrieved the corresponding process outputs. This verification step provides an assessment of the ML-predicted solutions against the underlying process simulation model without modifying the fixed flowsheet configuration.

Figure 10 compares the C-YIELD values predicted by the ML-based optimization platform with those recalculated using COCO for the 16 best optimization candidates selected from Figure 9. The comparison shows good agreement, with a mean absolute error (MAE) of 0.0085 and a Pearson correlation coefficient of 0.97. These results demonstrate that the ML-based optimization can identify operating conditions whose predicted performance is consistent with the reference process simulator.

Beyond the comparison of C-yield, CAAF retrieved additional COCO simulation outputs to evaluate the operating conditions and process performance of 16 selected candidates within the restricted optimization region (Table 2). The verified candidates are concentrated near the upper bounds of reactor temperature (485.9–486.0 K) and reactor length (6.99–7.00 m), with pressures ranging from 29.3 to 30.0 bar, splitter factors from 0.07 to 0.11, and CO molar flows from 47.6 to 50.5 mol/s. The corresponding methyl acetate production with C-YIELD varies between 0.89 and 0.92. These results indicate that the selected high-performance candidates occupy a relatively narrow operating region, particularly with respect to reactor temperature and length. Notably, the COCO-verified optimization results demonstrate a substantial improvement in C-Yield, reaching 0.92 compared with the maximum value of 0.81 obtained from the initial 2730 COCO process simulations (dataset 2). This represents an absolute increase of 10% in C-Yield, highlighting the effectiveness of the ML-based Bayesian optimization approach in identifying improved operating conditions beyond those explored in the initial simulation dataset.

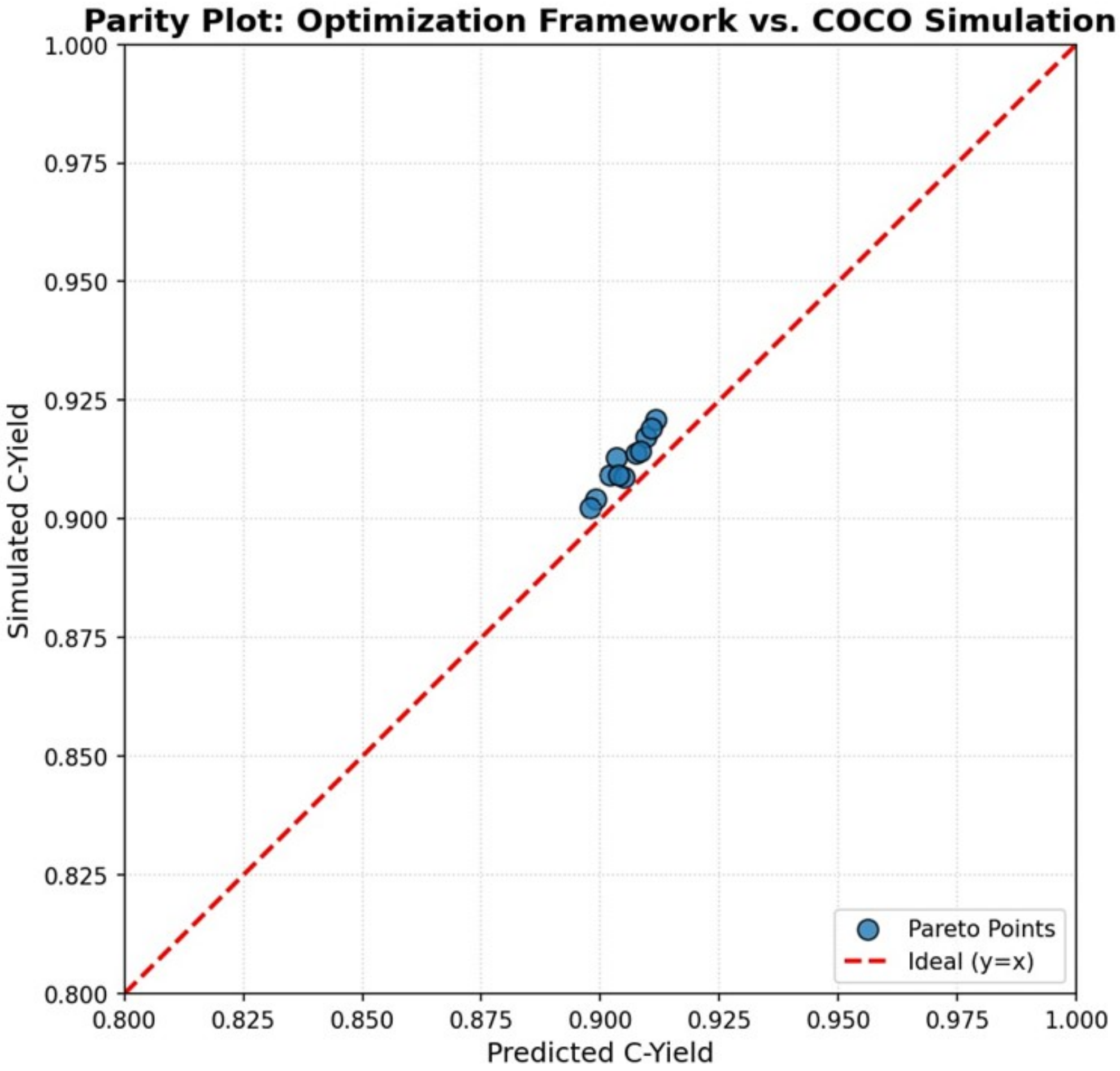

Figure 10. Comparison of C-YIELD predicted by the ML-based optimization platform and recalculated using COCO for selected optimization candidates. The reported MAE is 0.0085, with a Pearson correlation coefficient of 0.9713. The dashed line represents perfect agreement between predicted and simulated values.

An important outcome of this case study is the computational efficiency achieved through the integration of surrogate-based Bayesian optimization and rigorous process simulation within CAAF. Direct evaluation of all 4000 optimization candidates using COCO process simulations would require substantial computational resources and could take several weeks or months, depending on the computational setup and execution strategy. In contrast, the CAAF-coordinated workflow completed the Bayesian optimization search in approximately 2 minutes, followed by approximately 70 minutes for COCO-based verification of the selected optimized candidates (16 points). By using ML models to efficiently explore the multidimensional search space and limiting computationally expensive process simulations to selected high-performing solutions, the framework substantially reduces the number of rigorous simulation evaluations required during optimization. These results demonstrate the practical value of CAAF in integrating optimization, simulation-based verification, and engineering analysis into an efficient and interactive workflow, enabling engineers to identify and assess improved design and operating conditions within a substantially shorter evaluation cycle.

Table 2. Operating conditions and COCO-simulated process performance of Pareto-selected candidates from the restricted optimization region, retrieved and reported by CAAF (see stream and unit at figure 3).

| Point | Length | Temp | Pressure | Split Factor | CO Flow | MeAce | C-YIELD | Reboiler | Comp3 | Comp4 |
|---|---|---|---|---|---|---|---|---|---|---|
| 1 | 7,000 | 485,99 | 29,86 | 0,11 | 47,55 | 10436,9 | 0,921 | 1788,5 | 136,3 | 3,86 |
| 2 | 7,000 | 485,99 | 30,00 | 0,10 | 47,66 | 10437,1 | 0,920 | 1755,2 | 133,6 | 3,92 |
| 3 | 6,999 | 485,99 | 29,81 | 0,11 | 47,63 | 10433,7 | 0,920 | 1783,2 | 135,8 | 3,84 |
| 4 | 7,000 | 485,99 | 29,84 | 0,11 | 47,75 | 10431,4 | 0,919 | 1754,5 | 133,2 | 3,85 |
| 5 | 7,000 | 485,97 | 29,85 | 0,11 | 47,94 | 10426,7 | 0,917 | 1764,5 | 134,1 | 3,86 |
| 6 | 6,999 | 485,99 | 29,69 | 0,11 | 48,12 | 10417,7 | 0,915 | 1759,6 | 133,3 | 3,79 |
| 7 | 6,999 | 485,99 | 29,58 | 0,11 | 48,21 | 10413,6 | 0,914 | 1775,2 | 134,5 | 3,75 |
| 8 | 6,999 | 485,99 | 29,33 | 0,10 | 48,16 | 10404,6 | 0,914 | 1814,6 | 137,9 | 3,65 |
| 9 | 6,998 | 486,00 | 29,70 | 0,08 | 48,26 | 10402,6 | 0,913 | 1842,5 | 141,8 | 3,80 |
| 10 | 6,999 | 486,00 | 29,83 | 0,11 | 48,50 | 10412,7 | 0,912 | 1748,8 | 132,5 | 3,85 |
| 11 | 6,999 | 485,99 | 29,69 | 0,10 | 48,75 | 10401,1 | 0,909 | 1824,0 | 139,2 | 3,80 |
| 12 | 6,997 | 485,95 | 29,70 | 0,08 | 48,60 | 10388,9 | 0,909 | 1835,2 | 141,4 | 3,80 |
| 13 | 6,998 | 485,97 | 29,27 | 0,10 | 48,67 | 10389,7 | 0,909 | 1820,6 | 138,2 | 3,62 |
| 14 | 6,998 | 485,98 | 29,80 | 0,07 | 49,13 | 10372,7 | 0,904 | 1829,5 | 141,4 | 3,84 |
| 15 | 6,999 | 486,00 | 29,79 | 0,07 | 49,31 | 10366,9 | 0,902 | 1855,3 | 143,6 | 3,84 |
| 16 | 6,997 | 485,96 | 30,00 | 0,08 | 50,54 | 10332,1 | 0,891 | 1826,2 | 141,7 | 3,92 |

*Units: Length (reactor length) [m], Temp (Reactor temperature) [K], Press (reactor pressure) [Bar], CO (mole flow) [mol/s], MeAce (Methyl acetate Flow [kg/h], C-Yield (carbon conversion), Reboiler [kW], Comp3 (Compressor 3) [kW], Comp4 (compressor 4) [kW]*

Prompt 7 (table 2) was designed to retrieve additional process simulation outputs for selected candidate solutions within the Pareto-front region, particularly the MeAce production rates and energy requirements of individual process units under the corresponding operating conditions of the entire process. Through this prompt, CAAF extracted detailed simulation results from the established COCO flowsheet, enabling a more comprehensive evaluation of process performance beyond the primary optimization objectives of C-YIELD and C-SEC. Table 2 also provides additional information on equipment energy requirements, including reboiler and compressor duties. The reboiler accounts for a substantial energy demand, with its duty ranging from 1748.8 to 1855.3 kW, while the power requirements of compressors Comp3 and Comp4 range from 132.5 to 143.6 kW and from 3.6 to 3.9 kW, respectively. Although the candidate solutions exhibit relatively similar MeAce production rates, their equipment duties differ, illustrating the importance of considering downstream process performance when selecting operating conditions. For example, the candidate with the highest reported C-YIELD of 0.92 requires a reboiler duty of 1,788.5 kW (point 1), whereas another candidate with a comparable C-YIELD of 0,91 requires 1,842.5 kW (point 9). Such differences provide useful engineering information for evaluating alternative operating conditions beyond the primary yield objective.

## 4.2. Cognitive Agentic AI for Interactive Process Analysis and Optimization

The methyl acetate case study demonstrates the applicability of CAAF as an integrated environment for interactive process analysis and optimization, enabling engineers to coordinate heterogeneous engineering resources through successive natural-language prompts. Rather than introducing new process simulation models or optimization algorithms, the framework integrates established computational capabilities, including structured-data analysis, scientific knowledge retrieval, ML-assisted Bayesian optimization, and rigorous process simulation, within a unified investigation workflow. An important contribution of CAAF is its ability to maintain continuity across successive interactions, allowing engineers to progressively investigate complex multivariable relationships, interpret intermediate findings, refine optimization constraints, and request additional computational evaluations based on their engineering knowledge. This approach differs from executing individual engineering tasks independently, as the results of preceding analyses provide context for subsequent investigations without requiring users to manually coordinate multiple computational resources.

The computational performance of CAAF was evaluated through representative engineering tasks involving multivariable process analysis, Bayesian optimization, and direct process simulation. Comprehensive analysis of approximately 2730 DOE simulation cases involving five operating parameters required approximately 5 minutes, including data retrieval, statistical analysis, visualization, and interpretation supported by relevant scientific literature. Each multi-objective Bayesian optimization involving 4,000 iterations required approximately 2 minutes, comparable to the execution time of the optimization platform using the trained ML models without CAAF. For direct process simulation, executing COCO through CAAF introduced an additional overhead of approximately 1-3 seconds per simulation run compared with standalone COCO execution. Subsequent verification of 16 selected optimization candidates using COCO required approximately 70 minutes. These results demonstrate that CAAF can coordinate computationally demanding engineering tasks without substantially increasing the execution time of the underlying optimization and simulation tools. Although CAAF does not directly accelerate the numerical calculations, it provides an integrated interface for task formulation, computational execution, result interpretation, and subsequent investigation. The reported times characterize representative computational tasks rather than the complete end-to-end execution time of the entire interactive investigation.

The implementation of harness-engineering principles contributes to the structured coordination and reliability of CAAF operations. The architecture separates LLM-based reasoning, task planning, and resource orchestration from numerical calculations performed by specialized engineering tools. The Knowledge Guidance Layer provides domain-specific information to support the interpretation of engineering requests, while Memory maintains investigation context and intermediate results across successive interactions. The Verification module evaluates the consistency of executed tasks and returned outputs with the original engineering objectives. Furthermore, the framework incorporates execution monitoring, logging, and simulation failure-handling mechanisms to improve the transparency and traceability of computational workflows. When a process simulation fails to converge or encounters an execution error, CAAF reports the unsuccessful condition, excludes invalid numerical outputs, and continues evaluating subsequent simulation candidates. These mechanisms provide a structured environment for identifying execution problems and reviewing computational results, although they do not eliminate all potential LLM reasoning errors or uncertainties associated with the underlying engineering models.

The reliability and agent-level performance of CAAF were systematically investigated in a complementary study (Supplement information 2). The evaluation demonstrated a reproducibility rate of 93% across three repeated executions of identical prompts, an overall task execution accuracy of 81.5 ± 8.5 across the multi-agent framework, and a processing time of less than 120 seconds for agent-level planning and reasoning, excluding the execution time of external computational tools (with Gemma 4-31B open-weight LLM). These findings provide supporting evidence for the reliability and consistency of the underlying cognitive architecture. The present study builds on that evaluation by demonstrating how the framework can coordinate complementary engineering resources during an interactive process investigation involving multivariable analysis, optimization, and simulation-based verification.

Another important aspect of CAAF is the integration of human engineering expertise into the computational investigation. As demonstrated in Section 4.1, engineers can examine intermediate findings, identify promising operating regions, and introduce additional constraints based on technical knowledge and economic considerations. In the methyl acetate case study, the initial optimization results motivated the engineer to restrict reactor length, compressor pressure, and splitter factor before initiating a subsequent optimization. Selected optimization candidates were then evaluated using COCO to examine their predicted performance and equipment energy requirements. This interaction illustrates the practical value of Human-in-the-Loop (HITL) decision-making, where engineers retain control over investigation objectives, parameter constraints, and the selection of candidates for verification. In the present implementation, human interventions primarily occur between computational stages, while the individual numerical tasks are executed through the corresponding engineering tools.

The framework also provides opportunities to make advanced computational methods more accessible to process engineers who possess relevant domain knowledge but may not have extensive expertise in programming, database management, machine learning, or optimization algorithms. By translating natural-language requests into structured computational tasks, CAAF allows engineers to focus on interpreting process behavior, investigating parameter interactions, and evaluating alternative operating conditions rather than directly operating multiple specialized software environments. Nevertheless, the initial preparation of process simulation models, structured datasets, ML surrogates, and computational provider interfaces still requires specialist expertise. Therefore, the primary benefit of CAAF is to facilitate access to and coordinated use of established engineering capabilities rather than to replace the technical knowledge required to develop and validate these resources.

The modular and federated architecture further provides flexibility for integrating computational resources maintained by different engineering groups. Through dedicated Providers and defined interfaces, CAAF can connect separately managed process simulators, databases, optimization platforms, and scientific-document repositories while preserving the computational responsibilities of individual resources. This organization offers a potential pathway for extending the framework to other process technologies without fundamentally modifying its general cognitive architecture. However, the present study does not comprehensively investigate information security, confidentiality, or mechanisms for managing resource updates and version compatibility across independently maintained systems. These aspects require further development and systematic evaluation before deployment in more complex distributed engineering environments.

Future research should extend the evaluation of CAAF through systematic comparisons with alternative agent architectures and conventional engineering workflows, including end-to-end execution time, task completion accuracy, computational overhead, and the effectiveness of context reuse across successive interactions. The modular architecture also provides opportunities for integrating experimental measurements from laboratories, pilot-scale facilities, and industrial plants to support data-driven modelling, hybrid modelling, process analysis, and optimization. Such extensions would require reliable data acquisition, domain-specific model validation, and appropriate treatment of measurement noise, missing data, process variability, and model uncertainty. For applications involving industrial operations, additional safeguards and human approval mechanisms would be necessary to ensure that recommended conditions satisfy equipment limitations, operational constraints, and process safety requirements. Accordingly, CAAF should be regarded as an interactive engineering decision-support framework rather than an autonomous process control system.

Overall, the results demonstrate the potential of Cognitive Agentic AI to support a more accessible, integrated, and human-guided approach to process analysis and optimization. Its principal contribution lies in connecting human engineering expertise with heterogeneous computational resources through a structured and context-aware investigation workflow. By combining natural-language interaction, specialized engineering tools, and simulation-based verification, CAAF provides a foundation for developing more flexible and evidence-based computational environments for complex multivariable process investigations.

## 5. Conclusions

This study developed and demonstrated a Cognitive Agentic AI Framework (CAAF) for interactive process analysis and optimization by integrating heterogeneous engineering resources through natural-language interaction. The methyl acetate production case study demonstrated its applicability by connecting approximately 8,000 simulation cases, structured databases, scientific literature, ML-assisted Bayesian optimization, and process simulation within a continuous engineering workflow. Multivariable analysis provided insights into the relationships between operating parameters and process performance, while human-guided optimization enabled iterative refinement of the operating domain. Re-evaluation of 16 selected optimization candidates using COCO showed good agreement with ML predictions (MAE less than 1%). The highest COCO-verified C-Yield reached 0.921, compared with approximately 0.81 in the initial simulation dataset, representing an improvement of about 10 percentage points. Additional simulation results provided further insights into equipment, energy requirements and operating-condition trade-offs.

The main advantage of CAAF lies in making advanced process analysis and optimization more accessible to engineers by coordinating diverse engineering knowledge, data sources, and computational tools through natural-language prompts. Its harness-engineering architecture, combined with Human-in-the-Loop (HITL), enables automated computational workflows while allowing engineers to guide investigations, refine constraints, and evaluate intermediate results. Integrated verification, memory, and simulation failure-handling mechanisms further support the transparency, traceability, and reliability of model-based investigations.

Although demonstrated using simulation-generated data for methyl acetate production, the modular architecture of CAAF offers potential for extension to other process technologies, design-parameter optimization, and applications involving real pilot and real industrial plant data. Further validation and domain-specific integration will be required for these applications. Overall, CAAF provides a promising approach for combining human engineering expertise with Agentic AI to support flexible, interactive, and evidence-based process analysis and optimization.

## Data and Code Availability

Data, video demo of prompting on CAAF, the code of CAAF are publicly available at https://gitlab.sintef.no/trung.trinh/a-cognitive-agentic-ai-framework-for-interactive-process-analysis-and-optimization.

**Supplementary Information 1**

**Prompt 1:**
*Analyze the impact of operating parameters on the two primary objectives, carbon conversion yield and specific energy consumption (SEC) per kilogram of methyl acetate produced, with a focus on evidence obtained from the scientific literature repository and the DOE dataset stored in the SQL database.*

**Answer**
Based on the analysis of the DOE dataset and scientific literature, the impact of operating parameters on carbon conversion yield and specific energy consumption (SEC) for methyl acetate production is summarized as follows:
1. Impact on Carbon Conversion Yield

- Reactor Temperature (R-101_Temperature): This is the most critical driver. The DOE data shows a strong positive correlation (0.593), indicating that increasing temperature significantly enhances carbon efficiency. This aligns with reaction kinetics where higher temperatures typically increase the reaction rate of methanol carbonylation.
- CO Molar Flow (CO_molarFlow): Shows a notable positive impact (0.397). Literature supports this, noting that higher CO/DME ratios can increase conversion from 3% to over 99%, as CO is a primary reactant in the carbonylation process.
- Splitter Factors (Splitter_17_Split_factors_1): Exhibits a negative correlation (-0.277), suggesting that higher split factors in the separation unit may lead to increased carbon losses in the waste or recycle streams.

2. Impact on Specific Energy Consumption (SEC)

- Reactor Temperature & CO Molar Flow: Both show negative correlations (-0.381 and -0.387, respectively). This indicates that increasing these parameters reduces the energy required per kilogram of methyl acetate produced. This is likely due to improved kinetics and higher throughput, which spreads the fixed energy costs over a larger volume of product.
- K-101 Pressure (K_101_Pressure): Shows a positive correlation (0.318), meaning higher pressures increase the energy intensity. This is physically consistent with the increased work required by compressors (K-101) to achieve higher operating pressures.

Summary Table: Parameter Influence (generated from CAAF)

| Parameter | Carbon Efficiency | Specific Energy Consumption (SEC) | Overall Effect |
|---|---|---|---|
| Temperature | ↑ (Strong) | ↓ (Moderate) | Highly Beneficial |
| CO Molar Flow | ↑ (Moderate) | ↓ (Moderate) | Beneficial |
| Pressure | Neutral/Low | ↑ (Moderate) | Energy Penalty |
| Splitter Factor | ↓ (Moderate) | Neutral/Low | Detrimental to Yield |

Conclusion: To optimize the process for both high carbon yield and low energy consumption, the system should be operated at higher reactor temperatures and optimized CO molar flows, while keeping K-101 pressure at the minimum level required to maintain reaction stability and selectivity.

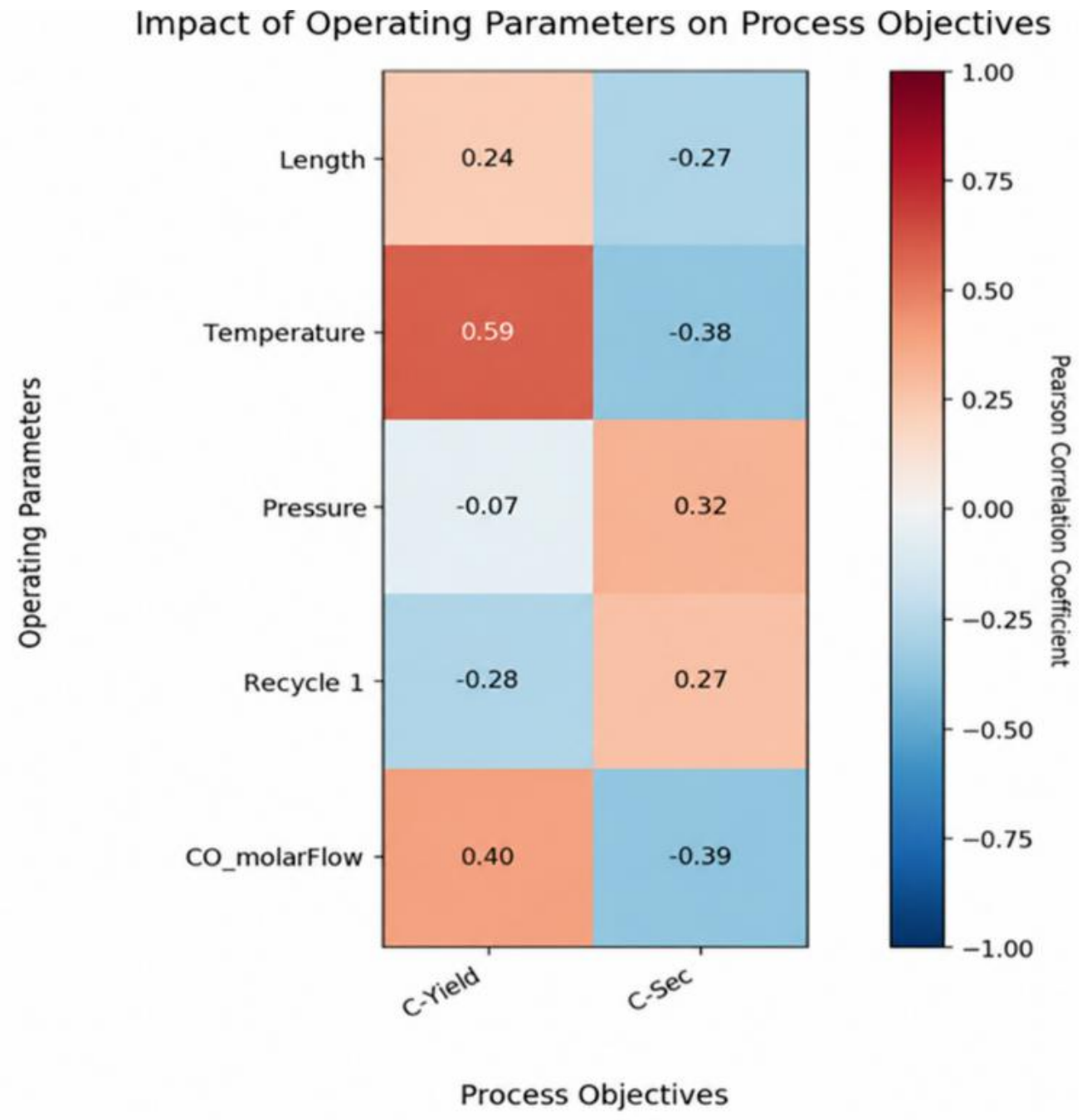

Figure 1: impact of operation parameters on process objectives generated from CAAF

**Prompt 2:**

Please visualize the relationship between the operating parameters and carbon conversion yield to gain a deeper understanding of the underlying multivariable effects. Use the data from the SQL database that was used to generate the DOE dataset and create the following plots:

- Plot three operating parameters together with carbon conversion yield in a single figure, where:
  - Y-axis: Carbon conversion yield (C-yield)
  - X-axis: One operating parameter and do not use splitter and pressure parameter.
  - Marker (symbol/category): Second operating parameter, with its values grouped into five categories (bins).
  - Color: Third operating parameter, with its values grouped into five categories (bins).
  - Include clear legends for both the color-coded parameter and the marker-coded parameter in every figure and include the corresponding value ranges for each bin.
- Use a jitter of 6 units for visualization.
- Generate plots for 10 plots, covering all possible combinations of 3 parameters out of 5.
- Use information the RAG system to explain the observed interaction effects between operating parameters on the resulting carbon conversion yield.

**Answer**

Based on the analysis of the DOE dataset (2,730 records) and the generated multivariable visualizations, the relationship between operating parameters and carbon conversion yield (C-yield) is characterized by the following interaction effects:

1. Primary Drivers of Carbon Conversion Yield

- R-101 Temperature (Strongest Positive Effect): Temperature is the dominant factor, showing a strong positive correlation (0.593). The data indicates a sharp increase in C-yield as temperature rises from 430K to 490K, particularly becoming more pronounced above 460K. This suggests that higher thermal energy is critical to overcoming the activation energy barriers of the catalyst.

- CO Molar Flow (Positive Effect): There is a significant positive correlation (0.397). Higher CO flow rates generally shift the yield upward, indicating that increased reactant availability enhances the overall carbon conversion efficiency.

2. Secondary and Inhibiting Factors
   - Splitter_17 Split Factor (Negative Effect): This parameter shows a negative correlation (-0.277), meaning that increasing the split factor tends to reduce the C-yield, likely by diverting reactants away from the primary reaction zone.
   - R-101 Length (Minor Positive Effect): Length has a modest positive correlation (0.238), suggesting that while a longer reactor increases residence time and thus yield, its impact is far less significant than temperature or flow rate.
   - K-101 Pressure (Negligible Effect): Pressure shows almost no correlation (-0.072), indicating that within the tested DOE range, the process is relatively insensitive to pressure changes.
3. Multivariable Interaction Insights
   - Synergistic Effect: The highest average C-yield (0.6291) is achieved through a synergy between 'Very High' Temperature and 'High' CO Molar Flow. This combination optimizes both the kinetic rate (via temperature) and the reactant concentration (via flow).
   - Optimal Operating Window: To maximize carbon efficiency, the process should be operated at: High R-101 Temperature High CO Molar Flow Low Splitter_17 Split Factors

These findings provide a clear roadmap for process optimization, highlighting that thermal management and reactant feed rates are the primary levers for improving carbon conversion efficiency.

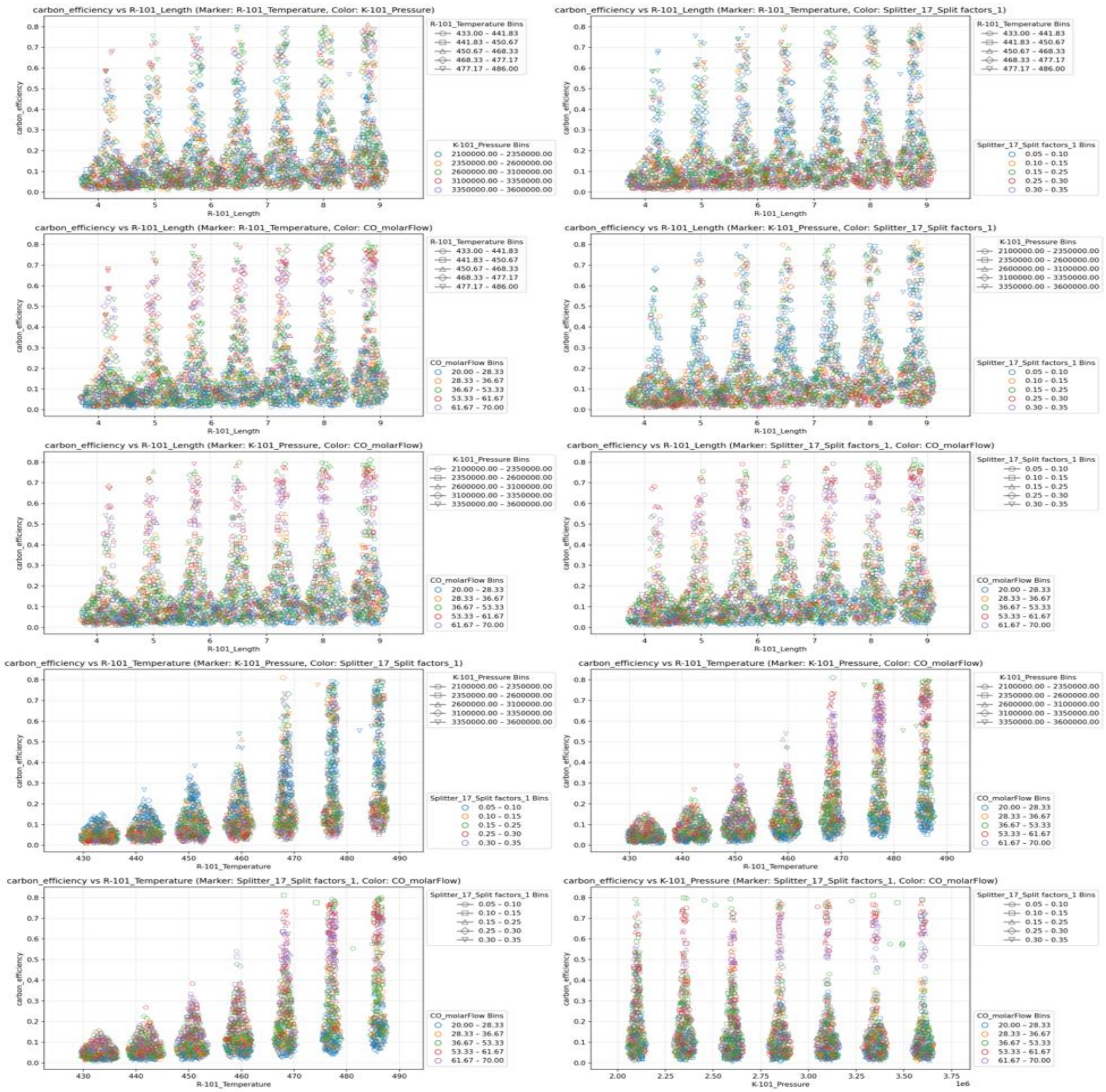

Figure 2: Multivariable relationship between operating variables and carbon conversion yield generated from CAAF

**Prompt 3:**

The pilot plant was operated under stable conditions for a long-term test from 2026 - 03 - 14, to 2026 - 05 - 15, using the same operating condition setting throughout the test period. Please plot the methyl acetate product stream together with main operation parameters with time and analyse the results.

**Answer**

Based on the analysis of the pilot plant data from March 14 to May 15, 2026, the process operated under generally stable conditions, although it exhibited high-frequency noise and occasional transient spikes.

Key Findings:

1. Product Performance: The Methyl Acetate (MeAce) mass flow averaged 4,693 kg/h with a Coefficient of Variation (CV) of 9.49%, indicating moderate variability despite the constant setpoints.
2. Operational Stability: The main control parameters remained highly consistent (all CV < 5%): R-101 Temperature: Mean = 476.93 K (CV = 0.67%) K-101 Pressure: Mean = 2.35 MPa (CV = 2.31%) CO Molar Flow: Mean = 44.99 (CV = 1.83%) R-101 Length: Mean = 8.00 (CV = 0.58%)

Analysis of Results:

- Temperature Correlation: A strong positive correlation (0.877) was identified between R-101 Temperature and MeAce mass flow. The data shows that sharp downward spikes in temperature directly coincide with dips in product flow, confirming that temperature fluctuations are the primary driver of product variability, consistent with reaction kinetics.
- Process Upsets: While the baseline was stable, the K-101 Pressure plot revealed three distinct high-pressure spikes (reaching ~3.0 MPa) and one sharp dip. These appear to be isolated equipment or process upsets rather than a systemic drift in catalyst activity or plant performance.

In summary, the plant maintained its target operating window, but the product output is sensitive to minor temperature fluctuations and occasional pressure transients.

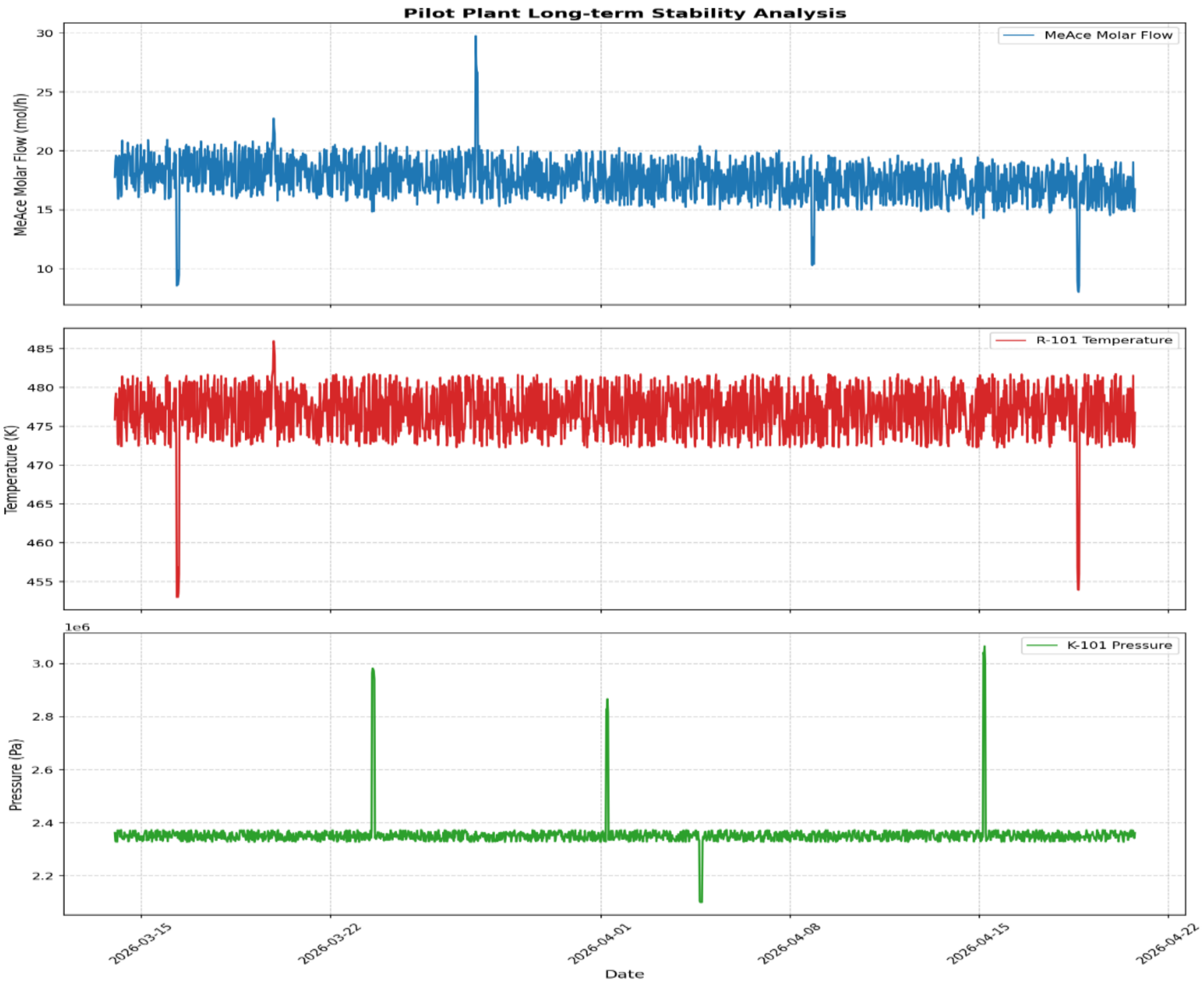

Figure 3: process parameters in stable conditions for a long-term test generated from CAAF

**Prompt 4:**

Using the process data collected during the long-term test period (2026-03-14 to 2026-05-15), generate time-series visualizations of
- the distillation column performance at 17 different tray locations, focusing on temperature profiles and heat duty profile
- methyl acetate mass flow
Analyze the influence of fluctuations in reactor operating conditions (CO flow rate, temperature, splitter ratio, and pressure) on Heat duty profile of the distillation column and methyl acetate mass flow

**Answer**
Based on the process data collected from 2026-03-14 to 2026-05-15, the analysis of the C-101 distillation column and reactor performance reveals the following:
1. Distillation Column Performance

- Temperature Profile: A clear thermal gradient is maintained across the 17 trays, ranging from approximately 301.57K at the top (Tray 1) to 386.04K at the bottom (Tray 17). This profile is consistent with the expected boiling point distribution for the separation of methyl acetate.
- Heat Duty Profile: The column operates with an average Cooler Duty (Top) of -20.57 MW and an average Reboiler Duty (Bottom) of 32.32 MW. The system generally maintains a steady state but is subject to periodic transient spikes.
- Methyl Acetate Mass Flow: The product flow averages 4,693 kg/h (standard deviation of 445.47 kg/h), showing synchronized fluctuations with the column's thermal duties.

2. Influence of Reactor Operating Conditions
Fluctuations in the reactor parameters propagate directly into the distillation section, causing the observed instabilities:

- Methyl Acetate Production: There is a strong positive correlation (0.877) between R-101 Temperature and methyl acetate mass flow. This indicates that reactor temperature is the primary driver of the ester production rate.
- Heat Duty Stability: The Reboiler Duty is strongly influenced by K-101 Pressure (correlation: 0.843). Pressure fluctuations in the column necessitate rapid adjustments in heat input to maintain the separation profile, leading to the observed spikes in duty.
- System Dynamics: High-frequency fluctuations in the Splitter ratio and Reactor Temperature act as the root causes for the synchronized 'spikes' seen across the distillation metrics, where increases in reboiler duty coincide with peaks in methyl acetate flow.

In summary, the distillation column's performance is highly sensitive to the upstream reactor's thermal and pressure stability, with R-101 temperature controlling the throughput and K-101 pressure driving the energy requirements for separation.

- Temperature (R-101): Operates around a mean of 476.93 K, with most values between 475 K and 482 K, though intermittent thermal disturbances caused spikes as low as 453 K and as high as 486 K.
- Pressure (K-101): Remains steady at approximately 2.35 MPa, with a few prominent positive spikes reaching up to 3.07 MPa and a sharp dip to 2.10 MPa.
- Split Factor (Splitter_17): Maintains a very stable setpoint of 0.10 (10%), with only negligible high-frequency precision fluctuations.
- CO Molar Flow: Highly stable at a mean of 44.99 mol/s, with notable outliers ranging from a drop to 32.72 mol/s to a spike of 54.99 mol/s.

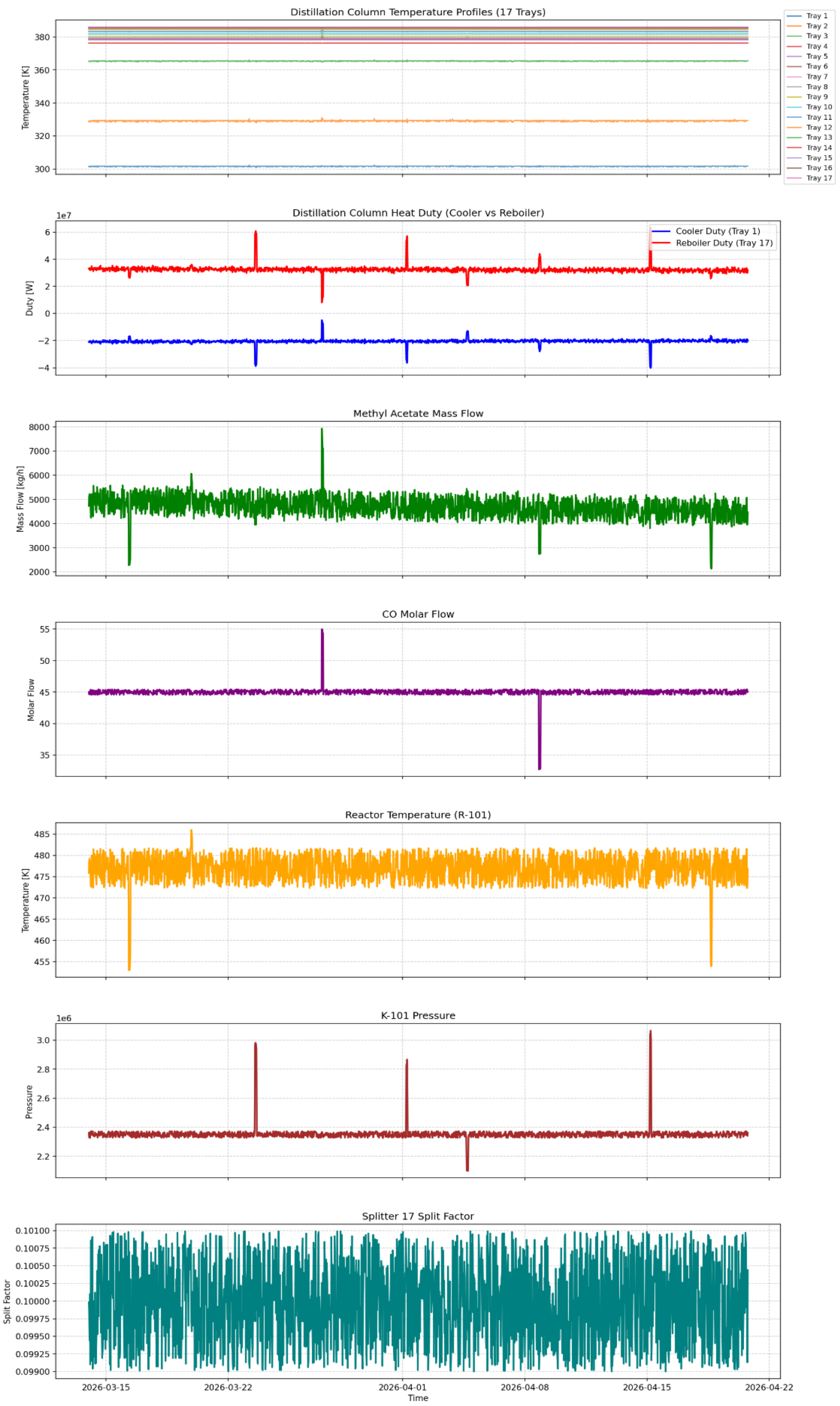

Figure 4: distillation parameters during stable conditions for a long-term test generated from CAAF

Key Observations from the Original Values:

**Prompt 5**

Please help me optimize two objectives: energy consumption cost (C-SEC) and carbon conversion efficiency (C-yield) using 4000 iterations within the following operating conditions:

- Reactor Length: 4.2 – 8.8 m
- Reactor Temperature: 433.0 – 486.0 K
- CO flow rate: 30 – 70 mole/s
- Compressor Pressure: 2.1 – 3.6 × $10^6$ Pa
- Split Factor: 0.05 – 0.35

Plot a graph showing the relationship between the two objectives during the optimization process where:

- The y-axis represents C-SEC on a logarithmic scale, with values less than 3
- The x-axis represents C-yield on a logarithmic scale
- Add the Pareto front to better illustrate the trade-off between the objectives Color indicates the iteration progression of the optimization process

Also, provide an explanation of the results.

**Answer**

The multi-objective optimization for Carbon Conversion Efficiency (C-yield) and Energy Consumption Cost (C-SEC) has been completed over 4000 iterations.

- Optimization Results Summary
- Trade-off Analysis: A strong negative correlation (-0.7495) was observed between C-SEC and C-yield. This confirms a fundamental process trade-off: achieving higher carbon efficiency typically requires a significant increase in energy expenditure.
- Performance Metrics:
    - Maximum C-yield: 0.9499
    - Minimum C-SEC: 0.0063
    - Average Performance: C-SEC of 0.2177 and C-yield of 0.8029
- Convergence: The optimization process successfully converged. Early iterations (represented in purple on the plot) explored a wide range of the objective space, while later iterations (yellow) clustered toward the optimal region (bottom-right of the log-log plot).
- Visualization Insights
- Pareto Front: The red line in the generated chart identifies 30 non-dominated solutions. These points represent the optimal balance where it is impossible to improve C-yield without increasing C-SEC.
- Diminishing Returns: The logarithmic scale reveals that as C-yield approaches its upper limit (~0.95), the energy cost (C-SEC) increases exponentially. This indicates a region of diminishing returns where marginal gains in efficiency become prohibitively expensive.
- Best Trial: Trial 3855 was identified as the most successful overall, balancing both objectives effectively.

In conclusion, the optimizer has successfully mapped the operational boundary of the plant. Depending on the current economic priority (energy saving vs. raw material efficiency), any point along the identified Pareto front can be selected as the target operating point.

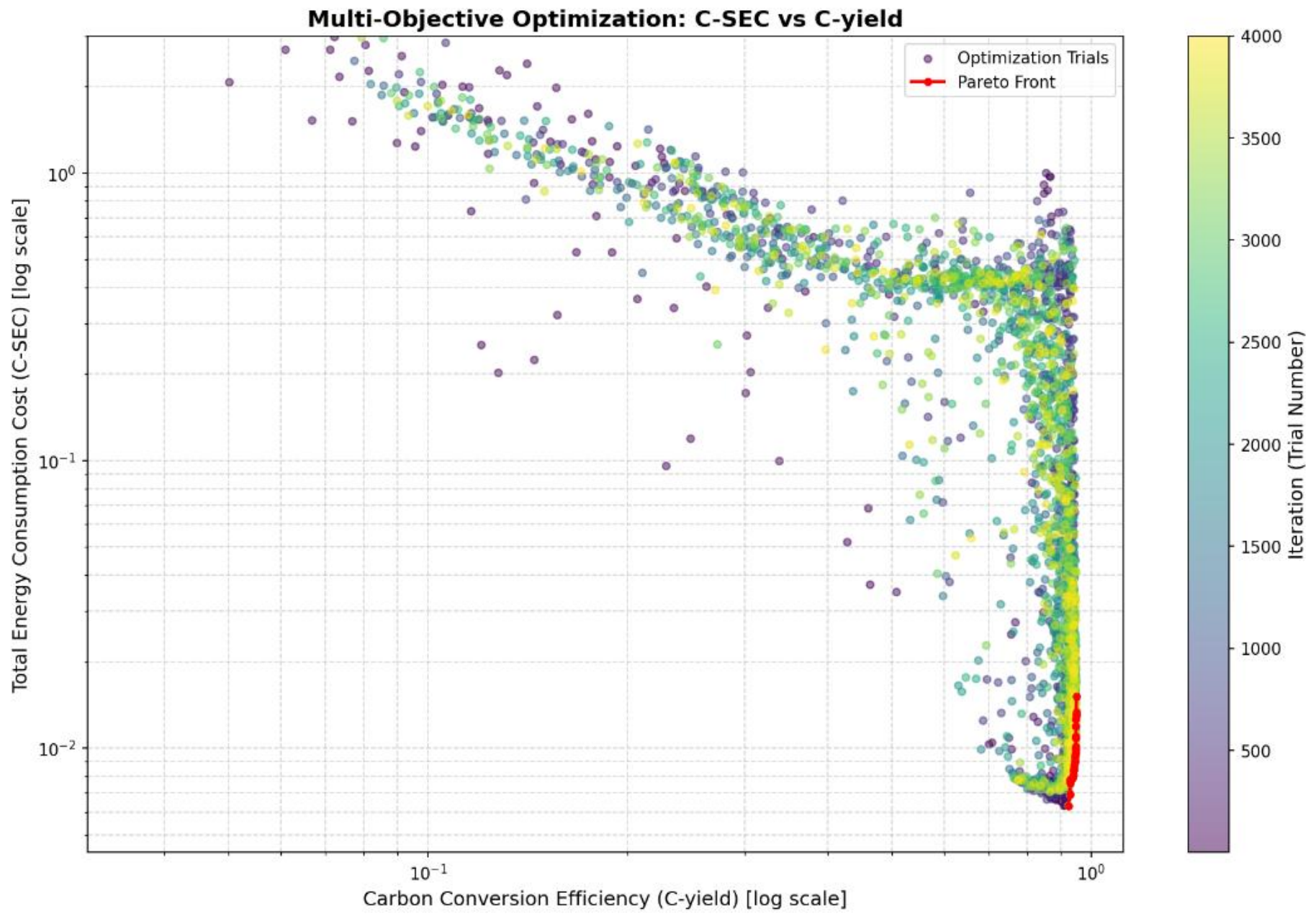

Figure 5: Multi-objective optimization results within the optimization space generated from CAAF

**Prompt 6**

Please create scatter plots for investigated parameters. In each plot:
- The x-axis represents C-yield.
- The y-axis represents energy consumption (on a logarithmic scale)
- Each operating parameter is represented by color I would like to examine how the operating conditions influence both C-yield and cost of energy consumption

**Answer**

Based on the multi-objective optimization results (4,000 iterations), the scatter plots analyzing the influence of operating parameters on Carbon Conversion Efficiency (C-yield) and Energy Consumption Cost (C-SEC) reveal the following insights:

- 1. Parameter Influence Analysis
- Reactor Length (R-101_Length): There is a strong positive correlation between reactor length and performance. Higher lengths (approaching 8.8m) are concentrated in the high-yield, low-cost region, indicating that increased residence time is critical for maximizing conversion efficiency while maintaining energy economy.
- Reactor Temperature (R-101_Temperature): Higher temperatures (up to 486K) shift the data points toward higher C-yields. This aligns with chemical kinetics, where increased temperature accelerates the reaction rate, though the most energy-efficient points are found at the upper end of the temperature range.
- Compressor Pressure (K-101_Pressure): Higher pressures (up to 3.6 MPa) generally correlate with higher C-yields. This is consistent with Le Chatelier's principle, as increased pressure favors the product side in this gas-phase reaction, effectively reducing the energy cost per unit of carbon converted.
- CO Molar Flow (CO_molarFlow): Lower flow rates (30–40 mol/s) are more prevalent in the high-efficiency, low-cost zone. Conversely, higher flow rates tend to decrease C-yield, likely due to the reduction in residence time within the reactor.

- Split Factor (Splitter_17): Lower split factors (~0.10) are strongly associated with the optimal region (high yield, low cost), whereas higher split factors (~0.30) correlate with lower yields and higher energy costs.
- 2. Summary of Optimal Operating Window

To maximize C-yield while minimizing C-SEC, the process should be operated toward the following conditions:

- Reactor Length: High (~8.8 m)
- Reactor Temperature: High (~486 K)
- Compressor Pressure: High (~3.6 MPa)
- CO Molar Flow: Low (~30 mol/s)
- Split Factor: Low (~0.10)

These results highlight a fundamental trade-off (correlation of -0.7495) between energy cost and carbon efficiency, but identify a clear path toward the Pareto front by optimizing these five key parameters.

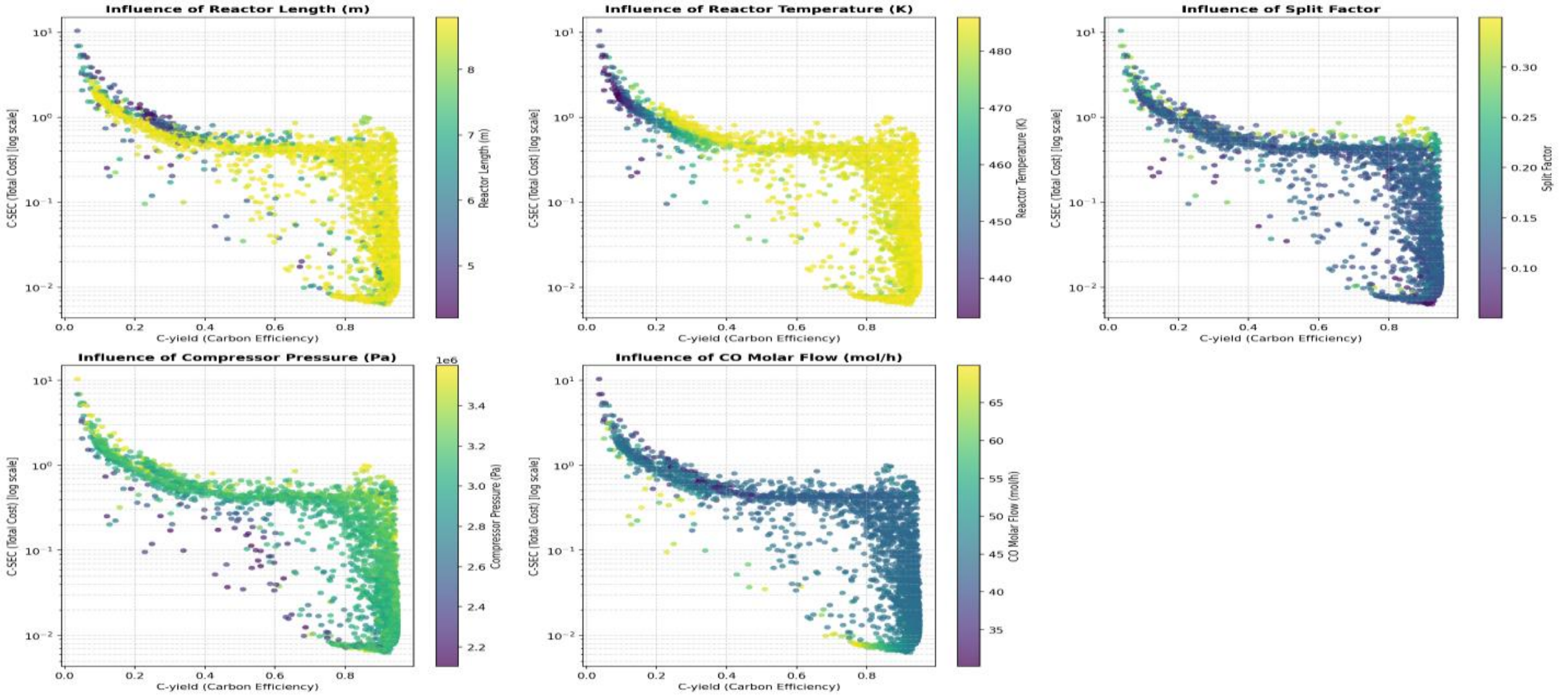

Figure 6: Distribution of operating variables within the optimization space generated from CAAF

**Prompt 7**

Based on the analysis of the DOE dataset stored in the database and the optimization results presented above, the findings suggest that the optimal carbon conversion yield may be achieved at pressures below *$3 \times 10^6$ Pa, reactor lengths shorter than 7 m, and splitter ratios below 11%.*

Please perform a new multi-objective optimization within this operating region with 4000 trials to determine the maximum achievable carbon conversion yield and the minimum specific energy consumption per kilogram of methyl acetate produced.

Plot a graph showing the relationship between the two objectives during the optimization process where:

• The y-axis represents C-SEC on a logarithmic scale, with values less than 0.01 and greater than 0.007

• The x-axis represents C-yield on a logarithmic scale greater than 0.85 and less than 0.92

• Add the Pareto front to better illustrate the trade-off between the objectives Color indicates the iteration progression of the optimization process

In addition, provide a preliminary assessment of the economic benefits of this operating strategy based on both OPEX (operating expenditure) and CAPEX (capital expenditure) considerations.

**Answer**

The multi-objective optimization within the constrained operating region (Compressor Pressure < 3 MPa, Reactor Length < 7 m, and Split Factor < 0.11) has been completed over 4,000 trials. The results demonstrate that high carbon efficiency can be maintained even with reduced equipment dimensions and lower operating pressures.

- 1. Optimization Performance
- Maximum Carbon Conversion Yield (C-yield): 0.9116
- Minimum Specific Energy Consumption (C-SEC): 0.007657
- Optimal Balanced Point: A yield of 0.9074 was achieved with a C-SEC of 0.008291.
- Pareto Front: 16 non-dominated points were identified, defining the optimal trade-off boundary. The optimization trajectory (visualized by color) shows strong convergence toward the bottom-right of the plot, indicating that the model successfully navigated the constrained space to find high-efficiency, low-energy configurations.
- 2. Economic Assessment

CAPEX (Capital Expenditure) Benefits:

- Reactor Sizing: By limiting the reactor length to < 7 m (average observed ~6.77 m), there is a direct reduction in the amount of alloy material required for the reactor vessel and a smaller physical footprint in the plant layout. This leads to lower procurement and installation costs.
- Compression Equipment: Operating at pressures below 3 MPa may allow for the use of lower-rated compressors or fewer compression stages, further reducing initial equipment costs.

OPEX (Operating Expenditure) Benefits:

- Energy Efficiency: The minimization of C-SEC to 0.007657 directly translates to lower utility costs. Specifically, the reduction in compressor pressure lowers electricity consumption, while the optimized split factor and reactor conditions minimize the steam and cooling water requirements per kg of methyl acetate produced.
- Resource Utilization: Achieving a C-yield of over 91% ensures maximum utilization of the CO feedstock, reducing raw material waste and lowering the cost of downstream separation/recycling.

In summary, this operating strategy successfully balances high productivity with significant reductions in both the initial investment and the ongoing cost of production.

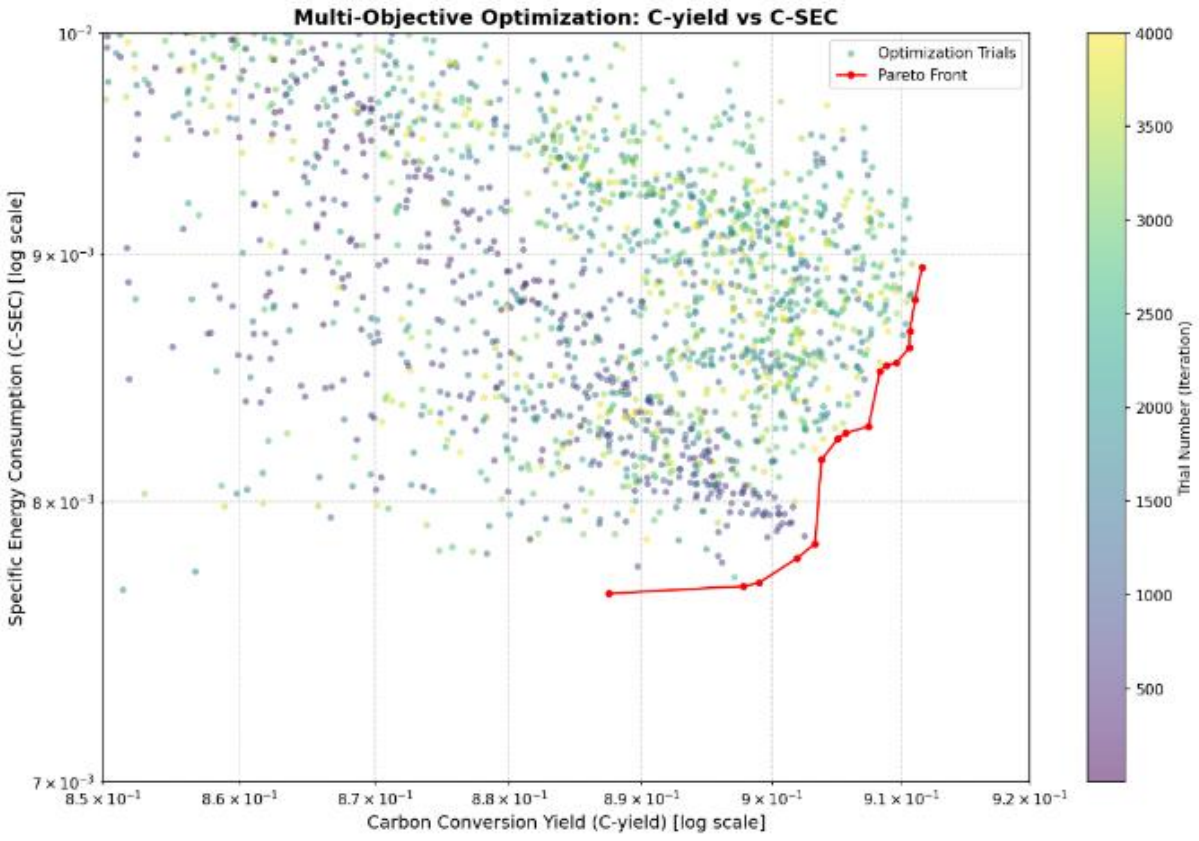

Figure 7: Multi-objective optimization results within the narrow optimization space generated from CAAF

**Prompt 8**

Please create scatter plots for investigated parameters. In each plot:

- The x-axis represents C-yield on a logarithmic scale greater than 0.85 and less than 0.92
- The y-axis represents C-SEC on a logarithmic scale, with values less than 0.01 and greater than 0.007

- Each operating parameter is represented by color I would like to examine how the operating conditions influence both C-yield and cost of energy consumption

**Answer**

The requested scatter plots have been generated to analyze the influence of operating parameters on the trade-off between Carbon Conversion Efficiency (C-yield) and Specific Energy Consumption (C-SEC), specifically focusing on the high-performance region (C-yield: 0.85–0.92; C-SEC: 0.007–0.01) on logarithmic scales.

- Key Parameter Insights:
- R-101 Temperature: This is a primary driver for productivity. Higher temperatures (approaching 486 K) are strongly correlated with the highest C-yields (up to 0.9116), though they typically increase the energy cost (C-SEC) due to higher thermal requirements.
- CO Molar Flow: There is a clear inverse relationship; lower molar flows are associated with both higher C-yields and lower C-SEC, suggesting that reducing the feed rate improves conversion efficiency and reduces specific energy demand.
- R-101 Length & K-101 Pressure: Both parameters show a trend where higher values (Length $\approx$ 7m, Pressure $\approx$ 3 MPa) push the system toward the Pareto front (bottom-right of the plots), indicating that increased residence time and pressure are essential for maximizing yield while maintaining efficiency.
- Splitter-17 Split Factor: Higher split factors (up to 0.11) generally correlate with improved C-yield, highlighting the importance of optimized separation/recycle ratios for carbon recovery.
- Optimization Summary:
- Maximum C-yield: 0.9116
- Minimum C-SEC: 0.007657
- Optimal Balanced Point: C-yield of 0.9074 at a C-SEC of 0.008291.

These results confirm that while maximizing yield requires higher temperatures and pressures, the most energy-efficient operation is achieved by balancing these with optimized molar flows and reactor dimensions.

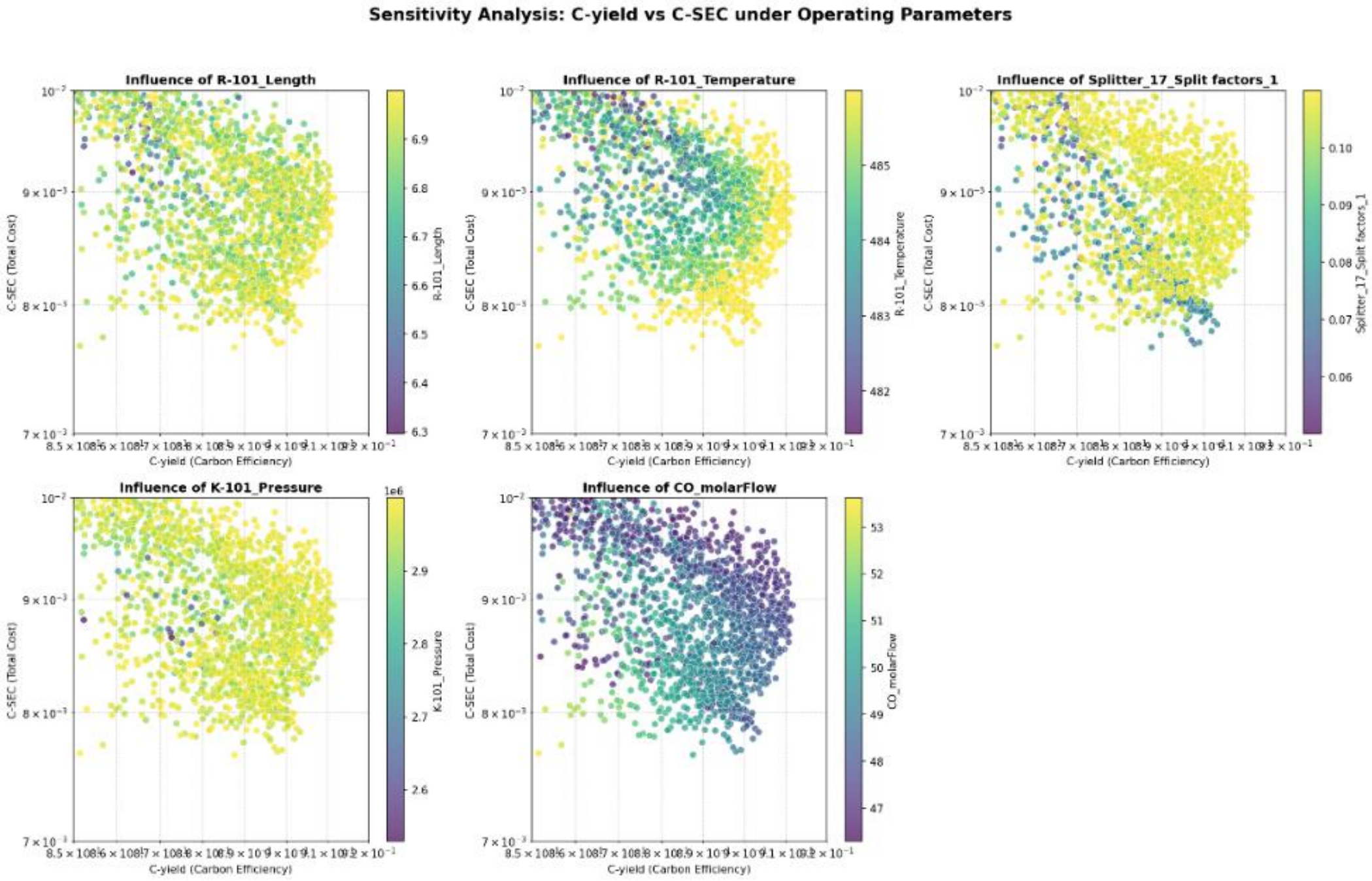

Figure 8: Distribution of operating variables within the narrow optimization space generated from CAAF

**Prompt 9**

Please rerun all optimized Pareto-front solutions generated by the previous optimization framework using the COCO process simulation model under the same operating conditions. Then, compare the results (C-YIELD) obtained from the optimization framework with those generated by the COCO process simulation and create a parity plot with the optimization framework results on the x-axis and the COCO simulation results on the y-axis. Include the reference line y=x and set both axes to range from 0.8 to 1.0. The comparison should clearly highlight any discrepancies in the predicted performance metrics and assess the accuracy of the optimization framework relative to the rigorous process simulation.

Answer

The validation of the Pareto-front solutions from the optimization framework was successfully completed by rerunning the 16 identified optimal points through the rigorous COCO process simulation model.

- Validation Results:
- Predictive Accuracy: The comparison between the predicted C-yield (from the optimization framework) and the simulated C-yield (from the COCO model) shows an exceptionally high degree of correlation.
- Parity Analysis: As visualized in the parity plot (chart_1783076239159.png), the data points are tightly clustered along the $y=x$ reference line, specifically within the high-performance range of 0.90 to 0.92.
- Observation: There is a consistent, albeit minimal, trend where the simulated values lie slightly above the reference line, suggesting that the optimization framework's surrogate model slightly under-predicts the actual carbon efficiency. However, this discrepancy is negligible for industrial application purposes.
- Conclusion:

The close alignment between the predicted and simulated results confirms that the optimization framework is highly accurate and reliable. The surrogate model effectively captures the non-linear dynamics of the process, ensuring that the identified Pareto-front solutions are both feasible and robust for implementation in the actual chemical plant.

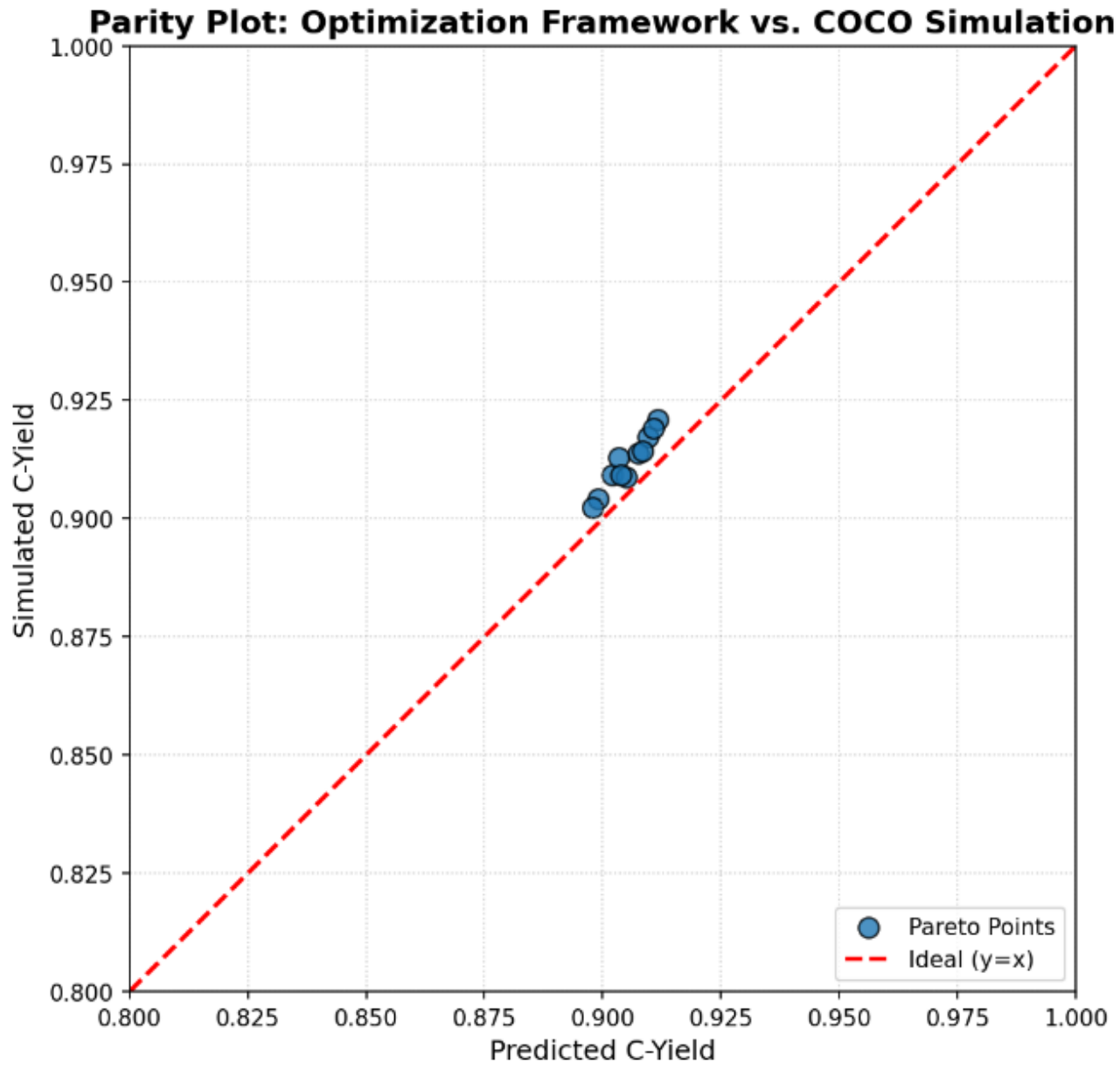

Figure 9: Comparison of C-Yield predicted by the optimization framework and rigorous COCO process simulation for Pareto-optimal solutions generated from CAAF

**Prompt 10**

Generate line charts for all Pareto-front solutions evaluated using the process simulation model and compare the following performance indicators with C-YIELD. Display all generated figures and provide a table for all Pareto-front solutions. The table should contain 10 columns: the five operating variables, C-YIELD, and the four performance indicators listed below.

Reboiler heat duty of the distillation column
Energy consumption of Compressor 3 (Comp3)
Energy consumption of Compressor 4 (Comp4)
Methyl acetate mass flow rate

**Answer**

The analysis of the Pareto-front solutions evaluated via the COCO process simulation model is complete. The results highlight the complex trade-offs between Carbon Efficiency (C-YIELD) and key energy and productivity indicators.

Performance Analysis

As visualized in the generated line charts (chart_1783078241934.png), the relationship between C-YIELD and the performance indicators is as follows:

- Methyl Acetate Mass Flow Rate: Exhibits a strong, positive linear correlation with C-YIELD. As efficiency increases from 0.891 to 0.921, the mass flow rate rises steadily from ~10,332 kg/h to ~10,437 kg/h, confirming that higher carbon efficiency directly translates to higher product yield.
- Reboiler Heat Duty: Shows a volatile relationship. While there is a general downward trend at higher efficiency levels, significant spikes occur (e.g., near 0.903 C-YIELD), indicating that the thermal demand is highly sensitive to the specific combination of operating variables.
- Compressor 3 Energy Demand: Generally decreases as C-YIELD increases (from ~141,800 kW down to ~132,500 kW), though the trend is non-monotonic with sharp fluctuations.
- Compressor 4 Energy Demand: Displays a complex, U-shaped pattern, reaching a minimum around 0.909 C-YIELD before increasing again at the highest efficiency levels, suggesting a non-linear energy penalty for achieving maximum conversion.

Pareto-Front Solutions Data Table

Below is the detailed data for the 16 Pareto-front solutions, sorted by C-YIELD in descending order:

| R-101 Length | R-101 Temp | K-101 Press | Split Factor | CO Flow | C-YIELD | Reboiler Duty | Comp3 Energy | Com4 Energy | MeAce Flow |
|---|---|---|---|---|---|---|---|---|---|
| 6.9998 | 485.99 | 2.986e6 | 0.1099 | 47.55 | 0.9210 | 1,788,582 | 136,398 | 3,866 | 10,436.9 |
| 6.9996 | 485.99 | 3.000e6 | 0.1025 | 47.66 | 0.9203 | 1,755,249 | 133,628 | 3,924 | 10,437.1 |
| 6.9990 | 485.99 | 2.981e6 | 0.1077 | 47.63 | 0.9202 | 1,783,235 | 135,841 | 3,848 | 10,433.7 |
| 6.9998 | 485.99 | 2.984e6 | 0.1063 | 47.75 | 0.9191 | 1,754,548 | 133,266 | 3,859 | 10,431.4 |
| 6.9996 | 485.97 | 2.985e6 | 0.1057 | 47.94 | 0.9173 | 1,764,589 | 134,181 | 3,865 | 10,426.7 |
| 6.9993 | 485.99 | 2.969e6 | 0.1084 | 48.12 | 0.9152 | 1,759,618 | 133,364 | 3,798 | 10,417.7 |
| 6.9992 | 485.99 | 2.958e6 | 0.1090 | 48.21 | 0.9142 | 1,775,283 | 134,580 | 3,757 | 10,413.6 |
| 6.9988 | 485.99 | 2.933e6 | 0.0990 | 48.16 | 0.9138 | 1,814,690 | 137,991 | 3,656 | 10,404.6 |
| 6.9981 | 486.00 | 2.970e6 | 0.0788 | 48.26 | 0.9129 | 1,842,599 | 141,854 | 3,804 | 10,402.6 |
| 6.9991 | 486.00 | 2.983e6 | 0.1054 | 48.50 | 0.9121 | 1,748,891 | 132,547 | 3,854 | 10,412.7 |
| 6.9993 | 485.99 | 2.969e6 | 0.1041 | 48.75 | 0.9093 | 1,824,098 | 139,257 | 3,800 | 10,401.1 |
| 6.9966 | 485.95 | 2.970e6 | 0.0773 | 48.60 | 0.9093 | 1,835,237 | 141,475 | 3,805 | 10,388.9 |
| 6.9976 | 485.97 | 2.927e6 | 0.1039 | 48.67 | 0.9089 | 1,820,674 | 138,216 | 3,629 | 10,389.7 |
| 6.9980 | 485.98 | 2.980e6 | 0.0743 | 49.13 | 0.9042 | 1,829,568 | 141,401 | 3,845 | 10,372.7 |
| 6.9990 | 486.00 | 2.979e6 | 0.0740 | 49.31 | 0.9024 | 1,855,351 | 143,652 | 3,840 | 10,366.9 |
| 6.9969 | 485.96 | 3.000e6 | 0.0754 | 50.54 | 0.8909 | 1,826,208 | 141,747 | 3,920 | 10,332.1 |

*(Note: Values are rounded for readability. Units: Length [m], Temp [K], Press [Pa], Flow [mol/h], Duty [kW], Energy [kW], MeAce Flow [kg/h])*

**Supplement Information 2.** Detailed Definitions of Evaluation Metrics

This appendix provides the formal definitions, and evaluation criteria for the supporting metrics used to characterize the underlying engineering capabilities of the proposed framework. These metrics are intended to provide component-level diagnostics for SQL retrieval, analytical execution, visualization, task understanding, tool selection, information retrieval, and response quality.

## 1. SQL Retrieval Metrics

### 1.1. Data Retrieval Accuracy

Data Retrieval Accuracy evaluates whether the SQL retrieval component obtains the appropriate information required by the downstream engineering workflow. The metric is assessed on a five-point Likert scale according to the completeness and correctness of the retrieved fields relative to the information required by the task.

The scoring criteria are:

- 5 - Excellent: All required columns or fields are retrieved correctly, with no unnecessary fields that materially affect the result.
- 4 - Very good: The required information is retrieved correctly, but a small number of additional columns are included that are not necessary for the task.
- 3 - Good: The required information is generally retrieved correctly, but a noticeable number of unnecessary columns are included.
- 2 - Poor: The query retrieves substantially more unnecessary information than required, reducing retrieval efficiency or introducing avoidable information into subsequent processing.
- 1 - Incorrect: One or more required columns are missing, or the retrieved columns are incorrect with respect to the task requirements.

### 1.2. Code Execution Error Rate

Code Execution Error Rate measures the proportion of generated SQL-related code executions that terminate with an execution error. A lower value indicates more reliable execution.

## 2. Data Analyst Metrics

### 2.1. Analytical Execution Accuracy

Analytical Execution Accuracy evaluates whether the Data Analyst produces an analytically correct result that adequately addresses the requested engineering analysis. The metric is assessed on a five-point Likert scale.

The scoring criteria are:

- 5 - Excellent: The analysis is fully correct, complete, and appropriately addresses the requested analytical objective.
- 4 - Very good: The analysis is correct and satisfies the task requirements, but lacks a relatively small amount of detail, depth, or analytical interpretation.
- 3 - Good: The main analytical result is correct, but the analysis is noticeably less detailed or insightful than expected.
- 2 - Poor: The analysis contains substantive shortcomings, such as incomplete reasoning, insufficient analysis, or partially incorrect conclusions.
- 1 - Incorrect: The analysis is incorrect or fails to satisfy the essential requirements of the task.

### 2.2. Code Execution Error Rate

Code Execution Error Rate measures the proportion of generated analytical code executions that fail during execution. This metric evaluates execution reliability independently of the quality of successfully completed analytical outputs.

## 3. Tool-Calling Quality

Tool-Calling Quality evaluates whether the Main Agent selects and invokes the appropriate engineering functionality with an appropriate task specification and parameterization. The assessment considers the selected tool or agent, the parameters supplied to it, and the task or information requirement passed through the call.

The metric is evaluated on a five-point Likert scale:

- 5 - Excellent: The correct tool or agent is selected, and the tool call contains the appropriate parameters, task specification, and required information.
- 4 - Very good: The selected tool or agent and the transmitted task information are appropriate, but the call provides an additional operation that is not strictly necessary because sufficient information or results were already available. The additional call is nevertheless relevant and does not constitute a materially redundant or incorrect operation.
- 3 - Moderate: The appropriate tool or agent is selected, but one or more parameters, task specifications, or transmitted requirements are imperfectly specified, with moderate impact on execution.
- 2 - Poor: The appropriate tool or agent is selected, but substantial errors occur in the parameters, task specification, or requirements transmitted to the component.
- 1 - Incorrect: An inappropriate tool or agent is selected, or the call is substantially irrelevant or incorrectly directed.

## 4. Visualizer Agent Metrics

### 4.1. Render Quality Score

Render Quality Score evaluates whether the generated visualization correctly represents the requested visualization objective and the information that should be communicated.

The metric is assessed on a five-point Likert scale:

- 5 - Excellent: The visualization fully satisfies the requested objective and is rendered correctly without substantive errors.
- 4 - Very good: The visualization satisfies the objective with only minor rendering or representation issues that do not materially affect interpretation.
- 3 - Moderate: The visualization addresses the requested objective only partially, with noticeable shortcomings in the representation.
- 2 - Poor: Only a limited portion of the requested visualization objective is satisfied.
- 1 - Incorrect: The visualization does not satisfy the requested objective.

### 4.2. Code Execution Error Rate

Code Execution Error Rate measures the proportion of visualization-generating code executions that fail. A lower value indicates more reliable visualization-code execution.

### 4.3. Visual Presentation Quality

Visual Presentation Quality evaluates the readability, organization, visual clarity, and overall presentation quality of the generated visualization. The metric is distinct from Render Accuracy: Render Accuracy evaluates whether the visualization communicates the requested information correctly, whereas Visual Presentation Quality evaluates how clearly and effectively that information is presented.

The metric is assessed on a five-point Likert scale:

- 5 - Excellent: The visualization is clear, well organized, visually coherent, appropriately labelled, and easy to interpret.
- 4 - Very good: The visualization is generally clear and well presented but contains minor presentation issues, such as small label overlaps or minor layout imperfections.
- 3 - Moderate: The visualization remains interpretable but has noticeable layout, readability, or aesthetic problems.
- 2 - Poor: The visualization has substantial presentation problems, including a difficult-to-follow layout, poor organization, or distracting visual elements.
- 1 - Unacceptable: The visualization is extremely difficult to interpret or provides little usable visual information.

## 5. Clarify Agent Metrics

### 5.1. Intent Disambiguation Score

Intent Disambiguation **Score** evaluates whether the Clarify Agent correctly identifies and explicitly represents the user's intended objective when the original request is ambiguous, incomplete, or contains multiple possible interpretations.

The metric is assessed on a five-point Likert scale:

- 5 - Excellent: The intended objective is correctly identified and expressed specifically and unambiguously.
- 4 - Very good: The intended objective is correctly identified, with only minor ambiguity or imprecision.
- 3 - Moderate: The general intent is identified, but the resulting interpretation remains somewhat vague or incomplete.
- 2 - Poor: The interpretation contains substantial ambiguity, confusion, or an incomplete representation of the intended objective.
- 1 - Incorrect: The inferred intent is inconsistent with the user's actual objective.

### 5.2. Task Classification Accuracy

Task Classification Accuracy measures whether the Clarify Agent assigns an input request to the correct predefined task category. A correct classification is assigned a score of 100%, whereas an incorrect classification is assigned a score of 0%.

## 6. Composite Semantic Score

The Composite Semantic Score evaluates the quality of an engineering response using three complementary dimensions: Accuracy, Groundedness, and Engineering Contextualization. Each dimension is scored on a five-point Likert scale, and the three scores are combined using the study-specific weighting scheme.

The composite score is defined as

$$Score = 0.5A + 0.3G + 0.2E$$

where:

- A is Accuracy;
- G is Groundedness; and
- E is Engineering Contextualization.

The resulting score remains on a five-point scale. Accuracy receives the largest weight because factual and numerical correctness is the primary requirement for industrial engineering decision support. Groundedness measures the extent to which the response is supported by available evidence, while Engineering Contextualization measures the degree to which the response appropriately interprets and explains the evidence in the relevant engineering context.

### 6.1. Accuracy

Accuracy evaluates the correctness of factual statements, numerical values, calculations, and logical conclusions with respect to the available evidence and the expected answer.

The scoring criteria are:

- 5 - Excellent: The response is fully correct with no substantive factual, numerical, or logical errors.
- 4 - Very good: The response is essentially correct but contains one or more minor errors that do not materially change the interpretation or conclusion.
- 3 - Moderate: The response contains several minor errors or one relatively noticeable error, while retaining substantial correct information.
- 2 - Poor: The response contains multiple substantive errors that affect part of the answer.
- 1 - Incorrect: The response contains serious errors or fails to provide a correct answer.

If a system execution does not produce an answer that can be evaluated, the Accuracy score is assigned as 1.

### 6.2. Groundedness

Groundedness evaluates whether the claims made in the generated response are supported by the retrieved evidence and available workflow information, without introducing unsupported factual claims or unjustified conclusions.

The scoring criteria are:

- 5 - Excellent: The response is fully supported by the available evidence and does not introduce unsupported or fabricated information.
- 4 - Very good: The response is well grounded but contains a minor unsupported inference or slight extrapolation.
- 3 - Moderate: The response contains some unsupported statements or inferences, but a substantial portion remains grounded in the available evidence.
- 2 - Poor: Several claims are insufficiently supported by the available evidence.
- 1 - Ungrounded: The response contains substantial unsupported claims, fabricated information, or lacks an evaluable evidence-based answer.

If a system execution does not produce an answer that can be evaluated, the Groundedness score is assigned as 1.

### 6.3. Engineering Contextualization

Engineering Contextualization evaluates whether the response appropriately interprets the available information within the relevant engineering context and provides useful domain-specific reasoning.

The scoring criteria are:

- 5 - Excellent: The response provides a complete, precise, and well-contextualized engineering interpretation that appropriately exploits the available information.
- 4 - Very good: The response is appropriately contextualized but omits a relatively small amount of useful engineering interpretation or detail.

- 3 - Moderate: The response is generally relevant to the engineering context but lacks sufficient detail or does not fully exploit the available information.
- 2 - Poor: The response provides limited engineering interpretation and is substantially generic.
- 1 - Unacceptable: The response is overly generic, excessively short, lacks meaningful engineering interpretation, or fails to provide useful domain-specific context.

If a system execution does not produce an answer that can be evaluated, the Engineering Contextualization score is assigned as 1.

## 7. Results

**Table 1.** Cross-LLM Component and Workflow Evaluation Results for Gemma-4 31B.

| Functional Module | Performance Metric | Run 1 | Run 2 | Run 3 | Mean ± SD |
|---|---|---|---|---|---|
| SQL Query Agent | Data Retrieval Accuracy (1-5) | 4.5 | 4.6 | 4.6 | 4.6 ± 0.1 |
| | Code Execution Error Rate (%) | 0.0 | 0.0 | 0.0 | 0.0 ± 0.0 |
| Data Analytics Engine | Analytical Execution Accuracy (1-5) | 4.2 | 4.5 | 4.5 | 4.4 ± 0.2 |
| | Code Execution Error Rate (%) | 22.2 | 4.2 | 0.0 | 8.8 ± 11.8 |
| Main Agent | Tool-Calling Precision (1-5) | 4.1 | 4.6 | 4.1 | 4.3 ± 0.3 |
| Visualizer Agent | Render Quality Score (1-5) | 4.4 | 4.0 | 4.1 | 4.2 ± 0.2 |
| | Code Execution Error Rate (%) | 10.3 | 7.5 | 5.0 | 7.6 ± 2.6 |
| | Visual Presentation Quality (1-5) | 4.5 | 4.6 | 4.6 | 4.6 ± 0.1 |
| Clarify Agent | Intent Disambiguation Score (1-5) | 5.0 | 4.8 | 4.8 | 4.9 ± 0.1 |
| | Task Classification Accuracy (%) | 91.7 | 100.0 | 100.0 | 97.2 ± 4.8 |
| RAG Module | Composite Semantic Score (1-5) | 4.8 | 5.0 | 5.0 | 4.9 ± 0.2 |
| Synthesizer Agent | Composite Semantic Score (1-5) | 4.5 | 4.3 | 4.3 | 4.4 ± 0.1 |
| Global Response Layer | Composite Semantic Score (1-5) | 4.1 | 4.0 | 4.0 | 4.0 ± 0.1 |